\documentclass[twocolumn,prl,aps,showpacs]{revtex4}
\usepackage{epsfig}

\begin{document}




\def\beq{\begin{equation}}
\def\eeq{\end{equation}}

\title{Magnetically-dominated jets inside collapsing stars
as a model for gamma-ray bursts and supernova explosions}

\author{Dmitri A. Uzdensky}
\email{uzdensky@astro.princeton.edu}
\affiliation{Dept. of Astrophysical Sciences, Princeton University,  
and Center for Magnetic Self-Organization (CMSO), Princeton, NJ 08544}

\author{Andrew I.\ MacFadyen}
\email{aim@ias.edu}
\affiliation{Dept. of Physics, New York University, and Institute 
for Advanced Study, Princeton, NJ 08540}

\date{March 7, 2007}

\begin{abstract}
It has been suggested that magnetic fields play a dynamically-important
role in core-collapse explosions of massive stars. In particular, they
may be important in the collapsar scenario for gamma-ray bursts (GRB),
where the central engine is a hyper-accreting black hole or a millisecond 
magnetar. The present paper is focussed on the magnetar scenario, with a 
specific emphasis on the interaction of the magnetar magnetosphere with 
the infalling stellar envelope. First, the ``Pulsar-in-a-Cavity'' problem 
is introduced as a paradigm for a magnetar inside a collapsing star. 
The basic set-up of this fundamental plasma-physics problem is described, 
outlining its main features, and simple estimates are derived for the 
evolution of the magnetic field. In the context of a collapsing star, 
it is proposed that, at first, the ram pressure of the infalling plasma 
acts to confine the magnetosphere,
enabling a gradual build-up of the magnetic pressure. At some point, the
growing magnetic pressure overtakes the (decreasing) ram pressure of the
gas, resulting in a magnetically-driven explosion. The explosion should 
be highly anisotropic, as the hoop-stress of the toroidal field, confined 
by the surrounding stellar matter, collimates the magnetically-dominated 
outflow into two beamed magnetic-tower jets. This creates a clean narrow 
channel for the escape of energy from the central engine through the star,
as required for GRBs. In addition, the delayed onset of the 
collimated-explosion phase can explain the production of large quantities
of Nickel-56, as suggested by the GRB-Supernova connection. Finally, the 
prospects for numerical simulations of this scenario are discussed.
\end{abstract}    
\vspace{0.1 in}


\maketitle

\newpage


\section {Introduction}
\label{sec-intro}


Long-duration gamma-ray bursts (GRBs) are believed to be strongly asymmetric 
explosions of massive stars. Some explosion asymmetry has also been 
inferred in core-collapse supernova (SN) explosions. The asymmetry 
is expected if the progenitor star is rapidly rotating, with the 
rotation axis providing a preferred direction for a jet-like outflow.

The leading theoretical model for long-duration GRBs based on rapid stellar 
rotation the {\it collapsar} model\cite{woosley-1993,paczynski-1998,
macfadyen-woosley-1999}. 
In this model the core of a massive star collapses to form a black hole. 
If the star is rapidly rotating, the overlying stellar material can form 
an accretion disk around the black hole. Because of differential
rotation in the disk, the magneto-rotational instability (MRI) is expected 
to develop ({\it e.g.}, Ref.~\cite{proga-2003}), providing angular momentum 
transport and dissipation of gravitational energy.  At the temperatures 
and densities ($T \sim 4 \times 10^{10}$ K, $\rho \sim 10^{10}$ g cm$^{-3}$) 
present in collapsar disks, neutrino emission cools the gas and the collapsing 
outer stellar core accretes at rates of $\sim 0.1\,M_{\odot}\,{\rm s}^{-1}$ 
for times $\gtrsim 10$~sec, long enough for the jet to break out of the star
and to power long GRBs. This accretion disk---black hole system thus acts 
as the central engine for the GRB. The energy source in this case is mainly 
accretion. Since most of this energy is released via neutrinos, most models 
have relied on neutrino luminosity to power the outflow. However, it has 
also been suggested that a significant amount of energy may be extracted 
magnetically and that strongly magnetized jets may play an active role in 
GRB explosions\cite{thompson-1994,meszaros-rees-1997,lee-2000,
vlahakis-konigl-2001,vanPutten-ostriker-2001,drenkhahn-spruit-2002,
lyutikov-blandford-2003,vanPutten-levinson-2003,proga-2003,lyutikov-2006,
mckinney-2005,um-2006a,giannios-spruit-2005,giannios-spruit-2006,
giannios-spruit-2007}).

In order for magnetic mechanisms to be viable, a fairly strong fields 
(of order~$10^{15}$~G) are required. Such fields, however, are expected 
to be generated by the MRI-driven turbulent dynamo in the disk during 
core-collapse\cite{proga-2003}.
Similar processes, resulting in similar field strengths, are also 
believed to be taking place in core-collapse supernovae explosions 
\cite{akiyama-2003,ardeljan-2005}.  
For example, Akiyama \& Wheeler have argued\cite{akiyama-2003} that 
the field may reach the level set by the equipartition with the MRI-driven 
turbulence, as strong as $10^{16-17}$~G. However, as they point out, this 
field is mostly toroidal. The large-scale {\it poloidal} magnetic field 
that is needed  may require a large-scale helical dynamo\cite{blackman-2006} 
and will probably be somewhat smaller than the toroidal field. Thus, we 
believe it is not unreasonable for the poloidal field at the disk surface 
to be a more modest $B_d\sim 10^{15}$~G (see also Ref.~\cite{proga-2003}).

In this paper we focus mostly on an alternative scenario, in which the 
central engine operating inside a collapsing star is not a black hole but 
a young millisecond magnetar --- a neutron star (NS) with a large-scale 
poloidal magnetic field of the order of $10^{15}$~G. Such a strong magnetic 
field can be produced by a convectively driven turbulent dynamo 
\cite{duncan-thompson-1992,thompson-duncan-1993} inside the young~NS,
or by the MRI-driven dynamo in the differentially rotating collapsing 
core\cite{akiyama-2003}.
An alternative possibility is that the progenitor core of about $10^4$~km 
has a magnetic field of order $10^9$~G, similar to the fields observed in 
some white dwarfs. When such a highly-magnetized core collapses into a 
neutron star of 10~km radius, flux freezing leads to the amplification 
of the magnetic field to $10^{15}$~G ({\it e.g.}, Ref.~\cite{um-2006a}).

We see that the typical values of the magnetic field strength, the rotation 
rate, and the size of the central engine in the classical collapsar scenario 
with an accreting black hole and in the millisecond-magnetar case are similar. 
The overall electro-magnetic luminosities should therefore be also comparable. 
And indeed, for a typical surface magnetic field $B_*=10^{15}$~G, 
a rotation rate $\Omega_*=10^4\ {\rm sec}^{-1}$, and a radius $R_*=10$~km,
the basic energetics and timescales make the millisecond magnetar 
a plausible candidate for a GRB central engine\cite{thompson-1994}. 
Ultimately, the energy source for the explosion is the rotational energy 
of the neutron star and the magnetic field acts as the agent that extracts 
this energy.  
In particular, the rotational energy of a millisecond-period neutron star 
is of order $E_{\rm rot}\simeq 5 \cdot 10^{52}\ {\rm erg}$, more than 
enough to drive a long-duration GRB.
The energy-extraction time scale, estimated using the usual pulsar luminosity 
formula $L_{\rm magn}\sim B_*^2\, R_*^6\, \Omega_*^4\, c^{-3}$, is then  
of order $100$~sec. 
Thus, the millisecond-magnetar central engine is essentially similar 
(with the magnetic field scaled up by three orders of magnitude and 
the timescale scaled down by six) to the Ostriker \& Gunn model\cite{og-1971} 
for powering supernovae by the spin-down magnetic power of a rapidly 
rotating pulsar. 

The idea of a millisecond magnetar as a central engine for gamma-ray bursts 
has been first proposed independently by Usov\cite{usov-1992} and by Duncan 
\& Thompson\cite{duncan-thompson-1992}. It has been further developed by 
a number of authors\cite{thompson-1994,yi-blackman-1998,nakamura-1998, 
spruit-1999,wheeler-2000,wheeler-2002,ruderman-2000,thompson-2004,um-2006b}. 
In addition, several magnetic explosion mechanisms have been explored in 
the supernova context\cite{leBlanc-wilson-1970,meier-1976,wheeler-2000,
wheeler-2002,akiyama-2003,thompson-2004,ardeljan-2005,blackman-2006,
shibata-2006,bucciantini-2006,metzger-2007,burrows-2007}). 
Our present paper is also devoted to investigating the millisecond-magnetar 
scenario for GRBs, but, importantly, viewed within the overall context of 
a collapsing star.

First we would like to stress that the plausible overall energetics
and timescales are, by themselves, not sufficient for making a good
GRB central engine model. The central engine also needs to be able 
to produce an energetic outflow that is (1) ultra-relativistic; 
(2) highly-collimated; and (3) baryon-free. Most of the previous studies 
have focused  mostly on the energetics and timescales, but not on 
the mechanisms for producing an outflow that satisfies these requirements 
(see, however, Refs.~\cite{wheeler-2000} and~\cite{bucciantini-2006} for 
a discussion of collimation). Also, most of these models, with the notable 
exception of Refs.~\cite{wheeler-2000} and~\cite{arons-2003}, have considered 
a magnetar in isolation, ignoring the effect of the surrounding stellar gas 
on shaping the outflow. 

One of the major points of our paper is that the infalling stellar gas 
is still present during the explosion and needs to be taken into account.
Thus, an important new element distinguishing our model from previous works 
is the consideration of the interaction between a newly-born magnetar and 
the stellar plasma in which it is initially embedded. Specifically, we argue 
that the pressure and inertia of the surrounding gas play a key role in 
regulating the magnetic extraction of rotational energy from the magnetar.
They also force the Poynting-flux-dominated outflow into two collimated 
jets, similar to Lynden-Bell's magnetic towers\cite{lb-1996,lb-2003,um-2006a}.

In Section~\ref{sec-stalled-shock} we discuss a specific situation relevant to 
the core collapse of a massive star: a cavity inside the stalled bounce-shock. 
The radius of such shock stays roughly constant allowing the magnetic fields 
in the cavity to grow. At the same time, both the ram pressure of the infalling
gas and the neutrino energy deposition inside the cavity decrease with time. 
We therefore argue that at some point, a fraction of a second after bounce, 
the magnetic field will start to dominate the force balance, leading to a 
magnetically-driven explosion. 

In order to illustrate these ideas, we introduce the ``Pulsar-in-a-Cavity'' 
problem as a basic-physics paradigm for this scenario. We describe this 
idealized problem in detail in~Section~\ref{sec-pulsar-cavity}. We first 
give a general description of the problem and its several versions. In 
Section~\ref{subsec-fixed-cavity}, we consider the simplest case of a 
force-free rotating magnetosphere inside a fixed rigid cavity. We demonstrate 
that differential rotation of the magnetic field lines is inevitably 
established inside the cavity, even though the pulsar itself is rotating 
uniformly; as a result, a strong toroidal magnetic field gradually builds up. 
We then study the long-term evolution of the field inside the cavity and show t
hat the magnetic luminosity increases with time. We also show that a massive, 
non-force-free plasma strip unavoidably arises in the equatorial plane 
beyond the light cylinder. We expect this phase to last until either the 
development of the kink instability modifies the situation or until the 
cavity walls yield to the internal magnetic stresses.
In Section~\ref{subsec-collimation} we discuss hoop-stress collimation and 
argue that external confinement and differential rotation are two important 
ingredients for collimating relativistic Poynting-flux dominated outflows. 
In either case, a magnetic tower forms. 

We investigate the propagation of a magnetic tower through the star in 
Section~\ref{sec-tower}. Section~\ref{subsec-lb-tower} is devoted to a 
general description of Lynden-Bell's magnetic tower model\cite{lb-1996}, 
whereas Sections~\ref{subsec-tower-star} --- \ref{subsec-numbers}
describe our modification of this model for the stellar environment. 
Specifically, we suggest that the external confining pressure invoked 
by Lynden-Bell's model is provided by the gas inside a hot cocoon behind 
a strong shock that the rapidly-growing magnetic tower drives into the 
unperturbed stellar envelope.

In~Section~\ref{sec-discussion}, we further explore some 
of the astrophysically-interesting aspects of our model. 
In Section~\ref{subsec-rel}, we consider the transition of the magnetic tower 
expansion to the relativistic regime and the final opening angle of the tower. 
In Section~\ref{subsec-small-scale}, we discuss a possible small-scale
substructure of the magnetic tower, represented as a ``train of plasmoids'',
that may be the outcome of the small-scale magnetic structure at the base 
of the outflow ({\it e.g.}, in the magnetized corona of the magnetar or of the 
accretion disk); it may also develop later as a result of MHD instabilities 
and flux conversion in the growing magnetic tower itself.
In Section~\ref{subsec-stability}, we discuss these stability issues, 
especially in regard to the Rayleigh--Taylor and kink instabilities. 
In Section~\ref{subsec-reconnection} we investigate the prospects for 
reconnection in the magnetar magnetosphere or in the magnetic tower, 
and argue that reconnection is ineffective in the dense environment 
deep inside the collapsing star. In Section~\ref{subsec-nickel}, we 
address an important issue of $^{56}$Ni production and argue that 
the two-phase nature of the explosion in our model is well-suited 
to explain a large amount of $^{56}$Ni inferred from observations. 
In Section~\ref{subsec-pulsar-kicks} we discuss the implications of 
our model for pulsar kicks. Finally, in~Section~\ref{subsec-numerical}, 
we suggest some directions for future numerical simulations of this 
problem. We draw our conclusions in~Section~\ref{sec-conclusions}.


\section{Magnetar inside a collapsing star: an outline of the general scenario}
\label{sec-stalled-shock}

\begin{figure} [h]
\centerline
{\psfig{file=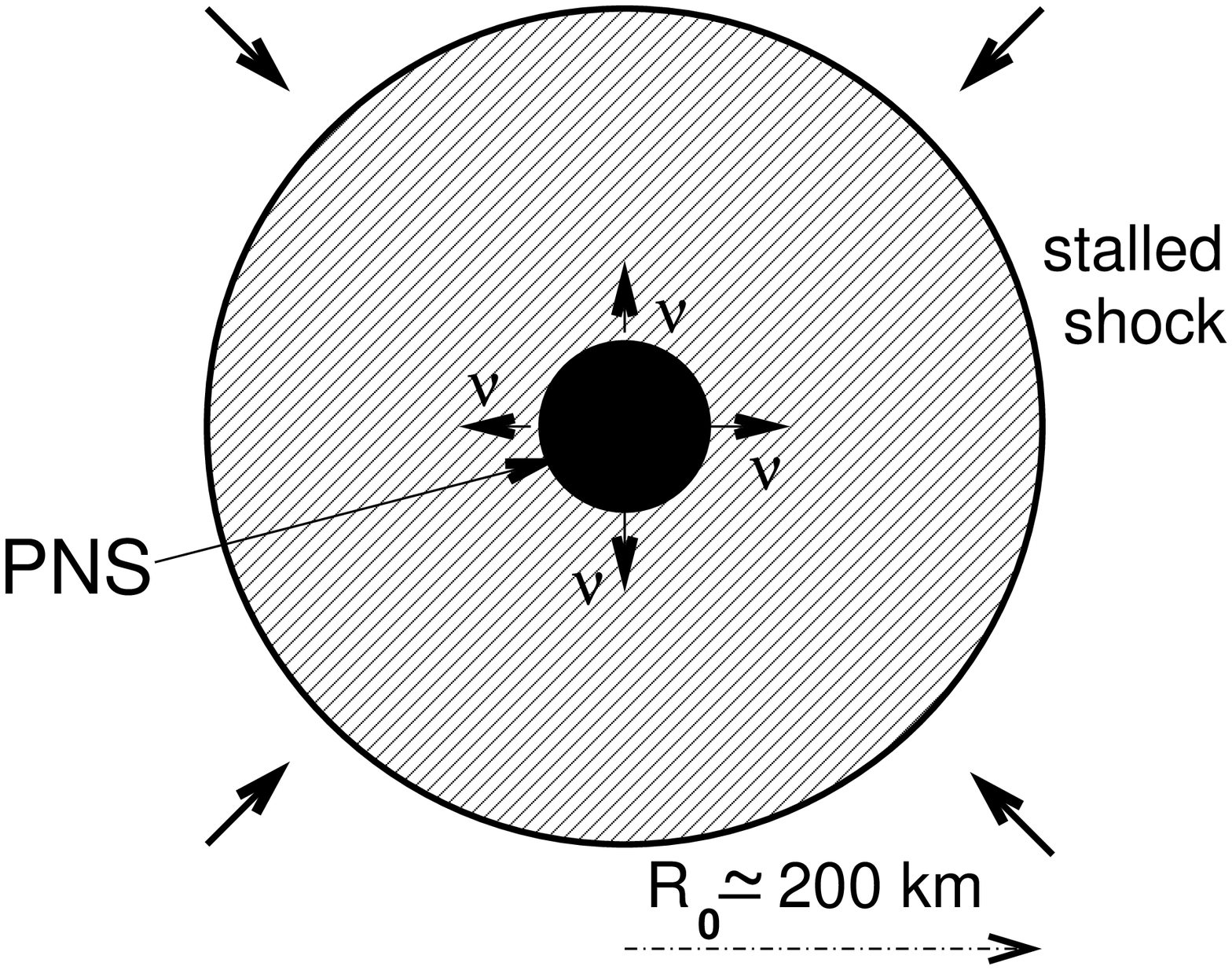,width=5in}}
\caption{Stalled shock phase of core-collapse explosion.}
\label{fig-stalled-shock}
\end{figure}

Numerous studies of core-collapse supernovae have shown that, as the core 
of a massive star collapses into a proto-neutron star (PNS), a bounce shock 
is launched back into the star but quickly stalls at about 200~km (see, 
{\it e.g.}, Refs.~\cite{woosley-weaver-1986,bethe-1990}). 
The explosion then enters a long ($\sim$ 1~sec) phase 
(see Fig.~\ref{fig-stalled-shock}) during which the shock 
is quasi-stationary as determined by the balance between the ram 
pressure of the infalling material, which tends to quench the shock, 
and the thermal pressure of the post-shocked gas, supported by the continuous 
neutrino heating. Eventually, if neutrinos win, the shock engulfs the entire 
star and one gets a successful SN explosion. If they lose, the shock dies 
and the PNS collapses into a black hole that subsequently swallows the rest 
of the star, without a~SN.

\begin{figure} [h]
\centerline
{\psfig{file=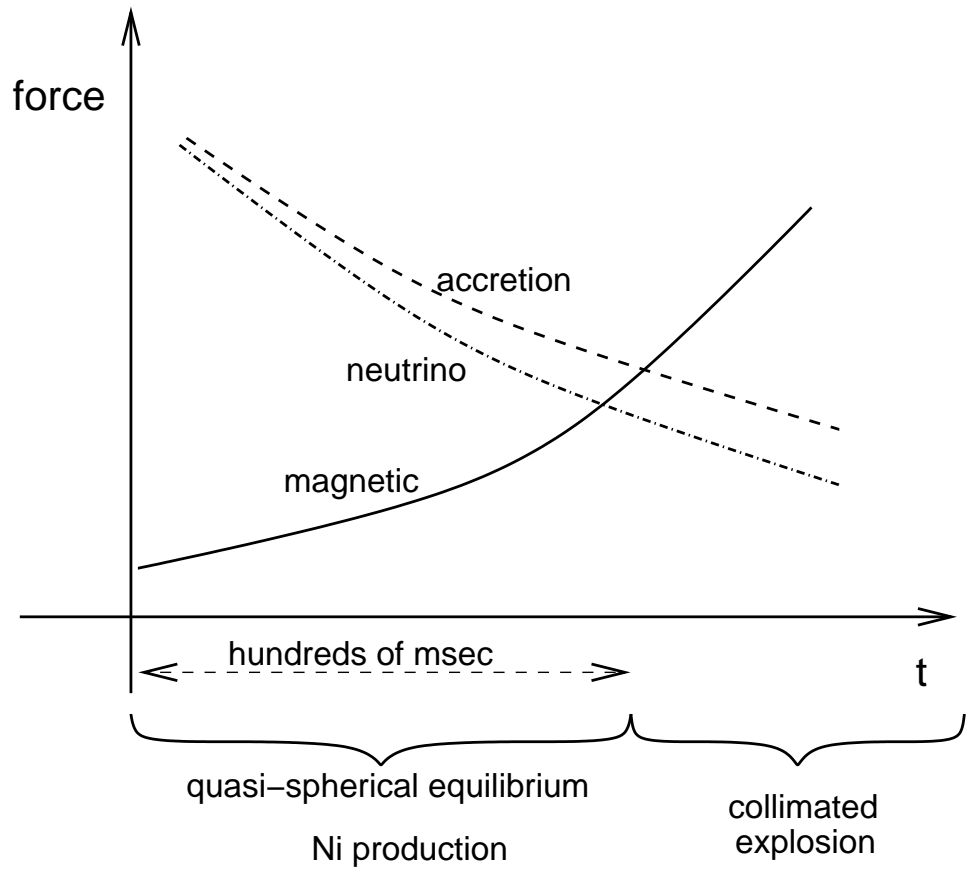,width=5in}}
\caption{Schematic time evolution of the main three forces responsible 
for the stalled-shock force balance.}
\label{fig-scenario}
\end{figure}

In our model, we add a third dynamical component to this picture --- 
the magnetic field. The magnetic pressure is pushing out, helping the 
explosion, as is the thermal pressure of the neutrino-heated gas. Our 
main idea is that these two outward forces evolve differently with time, 
and thus the explosion may be a two-stage process. In particular, we 
suggest that the magnetic pressure is not important during the first 
few hundreds of msec of the stalled-shock phase. However, during this 
time the magnetar makes several hundred revolutions. This results in 
a great amplification of the toroidal magnetic flux by the differential 
rotation (see~Sec.~\ref{subsec-fixed-cavity}), whereas both the neutrino 
energy deposition and the accretion rate gradually decline (see Fig.~\ref
{fig-scenario}).
For example, assuming $R_0=3 R_{\rm LC}= 10 R_*=100$~km, and $B_*=10^{15}$~G, 
the entire cavity is filled (see~Sec.~\ref{subsec-fixed-cavity}) with 
$3\cdot 10^{14}$~G fields after about 100~turns (0.1~sec), corresponding 
to the magnetic pressure of $4\cdot 10^{27}$ erg/cm$^3$. 
The ram pressure of the infalling stellar material at $r=R_0$, compressing 
the magnetosphere, can be estimated roughly as $P_{\rm ram} \sim \dot{M} 
v_{\rm ff}/4\pi R_0^2 \simeq 8 \cdot 10^{27}\, \dot{M}_0\, M_0^{1/2}\, 
R_{0,7}^{-5/2} {\rm erg\ cm^{-3}}$
where $v_{\rm ff}=(2GM/R_0)^{1/2}\simeq 5\cdot 10^9\, M_0^{1/2}\, 
R_{0,7}^{-1/2}\, {\rm cm/sec}$ is the free-fall velocity at radius~$R_0$,
and~$M_0$ and~$\dot{M}_0$ are the mass within~$R_0$ and the accretion rate 
at this radius, expressed in units of~$M_\odot$ and~$M_\odot/{\rm sec}$, 
respectively.
This shows that, after a delay of a few hundreds of milliseconds, 
the magnetic pressure inevitably becomes an important driving force 
and may lead to a successful explosion.

To summarize our picture, the ram pressure of the accreting material 
provides a nurturing womb in which the baby magnetic field grows, 
until it is finally strong enough to break out. Neutrino energy 
deposition plays an important role during this gestation period, 
as it prevents the magnetosphere from being completely squashed 
by the accreting gas.
Finally, if the above picture is correct and the explosion does 
become magnetically-driven, then the hoop-stress mechanism makes 
it highly collimated, thus satisfying one of the key necessary 
conditions for GRB (see~Sec.~\ref{subsec-collimation}). Note that 
this jet is driven by the magnetar-level ({\it i.e.}, $\sim10^{15}$~G) 
field and is thus much stronger and faster than the LeBlanc-Wilson 
jet\cite{leBlanc-wilson-1970} that may have been launched a few seconds 
earlier, during the core-collapse process\cite{wheeler-2000}.


\section{The Pulsar-in-a-Cavity Problem}
\label{sec-pulsar-cavity}

\begin{figure} [h]
\centerline
{\psfig{file=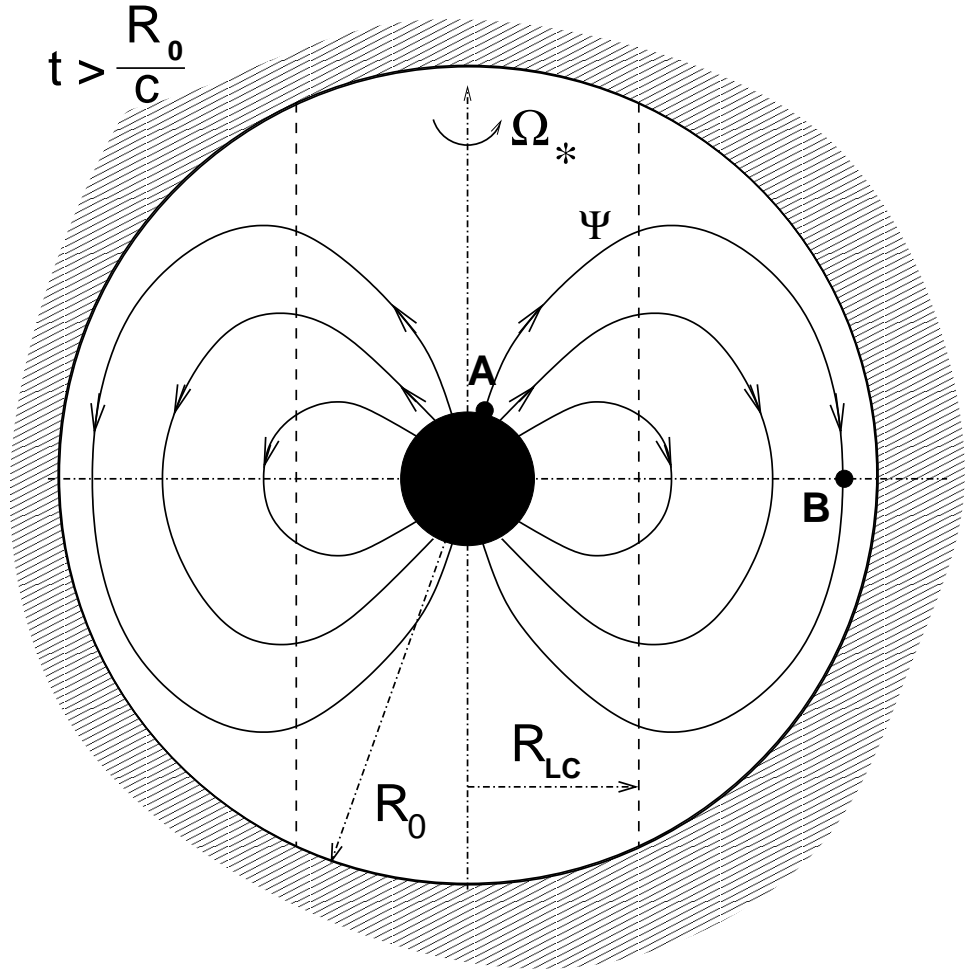,width=5in}}
\caption{Aligned pulsar inside an infinitely-conducting 
spherical cavity of radius~$R_0$. The vertical dashed lines represent 
the pulsar's light cylinder of radius $R_{\rm LC}<R_0$. 
After a time of order the
light-crossing time~$R_0/c$, the poloidal field lines outside 
the light cylinder expand somewhat but still remain confined within 
the cavity. Because the toroidal magnetic field has to vanish 
at the equatorial midplane due to reflection symmetry, the field 
lines there cannot corotate with the star, $\Omega_B<c/R_B<\Omega_*$.
As a result, differential rotation is established in both hemispheres,
$\Delta\Omega=\Omega_*-\Omega_B\simeq\Omega_*$ (for $R_B\gg R_{\rm LC}$),
which leads to continuous generation of toroidal magnetic flux.}
\label{fig-magnetar-2}
\end{figure}

As a first step in trying to understand how a millisecond-magnetar 
central engine works in the collapsar context, let us consider the 
following basic plasma-physics problem: an axisymmetric pulsar inside 
a conducting cavity with a low-density plasma (see~Fig.~\ref{fig-magnetar-2}). 
Specifically, we consider the cavity's radius~$R_0$ to be is much larger 
than the pulsar light-cylinder radius~$R_{\rm LC}$. We call this idealized 
problem the {\it Pulsar-in-a-Cavity problem}\cite{um-2006a,um-2006b}.
It is a modification of the Goldreich \& Julian model for an isolated 
pulsar's magnetosphere\cite{gj-1969}; it has direct connections to the 
model proposed by Ostriker \& Gunn for powering longer-term supernova 
lightcurves\cite{og-1971} and it is also related to the models considered 
by Kardashev\cite{kardashev-1970} and by Illarionov \& 
Sunyaev\cite{illarionov-sunyaev-1975}.

Note that the behavior of Pulsar-in-a-Cavity depends on the cavity properties. 
For definiteness, we assume that both the walls and the plasma inside the 
cavity are perfect conductors. We also assume that all the field lines close 
back to the pulsar inside the cavity, keeping in mind that the magnetic field
had been generated inside the NS and then emerged through its surface.

At the same time, we are dealing with a whole family of problems 
distinguished by the assumed mechanical properties of the cavity.
For simplicity, we shall concentrate on the case of a spherical 
cavity with rigid walls (see Sec.~\ref{subsec-fixed-cavity}), 
which may correspond to early stages of the system's evolution. 
In the future we plan to consider more realistic but also more 
complicated cases where the shape and the size of the cavity are 
not fixed but instead are governed by a balance with an external 
pressure (similar to Lynden-Bell's magnetic tower model\cite{lb-1996}),
and, finally, will consider a young pulsar in a fully dynamic 
environment of a collapsing star (c.f., Ref.~\cite{og-1971}).

Each of these versions of our pulsar-in-a-cavity problem will provide 
important insights into the workings of a millisecond magnetar inside 
a collapsing star. They will probably require numerical simulations
using relativistic force-free  or  relativistic MHD codes (see~Sec.~\ref
{subsec-numerical}). 
To set the stage for these numerical studies, we will, in this section, 
qualitatively discuss a plausible physical picture of the system's evolution.

To investigate the interaction between the central magnetar and 
the surrounding stellar material, a full magnetohydrodynamic (MHD) 
description that includes plasma pressure and inertial effects will 
eventually be required. This is especially relevant if there is a 
strong wind driven off the~PNS by neutrinos and/or by the 
magneto-centrifugal mechanism ({\it e.g.}, Refs.~\cite{thompson-2004, 
bucciantini-2006,metzger-2007}). 
Of particular interest is the confinement of 
the expanding magnetosphere by the surrounding plasma and the dynamical 
response of the star to the expanding magnetosphere at its center. 
For simplicity, however, here we shall limit ourselves to 
the relativistic force-free case in which electromagnetic forces 
dominate the dynamics almost everywhere inside the cavity (except 
at the the equatorial plane outside the light cylinder --- see below).
While it is not realistic in the central region of a collapsing star, 
the force-free description may nonetheless reflect some essential physics, 
especially for late phases of the evolution.

We would like to remark that the ideal-MHD assumption is well justified 
due to the very large plasma densities and temperatures. Specifically, 
the high plasma density ensures that the plasma is highly collisional 
and hence is well described by resistive MHD; other non-ideal terms in 
generalized Ohm's law are unimportant. On the other hand, the resistivity 
is actually small in absolute terms ($Re_m \gg 1$). All this makes ideal MHD 
a good approximation in the environment of a collapsing star\cite{um-2006a}. 
Therefore, in this paper we shall ignore any non-ideal effects, leaving 
them for a future study.

As a final point, in contrast with the isolated pulsar, which is usually 
regarded as stationary, our pulsar-in-a-cavity problem is intrinsically 
time-dependent.


\subsection{Pulsar in a Fixed Spherical Cavity}
\label{subsec-fixed-cavity}

In this section we consider the case of a spherical cavity 
whose radius, $R_0$, is kept constant (or is changing slowly).
We are interested in the evolution of an initially dipole-like 
magnetosphere after the pulsar is spun up suddenly.
In the isolated pulsar case, the field lines extending beyond 
the light cylinder are swept backwards and open\cite{gj-1969}; 
there is no feedback of the outer region on the inner magnetosphere. 
In our case, however, the entire magnetosphere is contained inside 
the cavity and hence remain closed at all times. This is a very 
important difference between the two cases.


The first thing to note is that a {\it differential rotation} 
is established in the magnetosphere. Indeed, consider a field 
line~$\Psi$ that extends beyond the light cylinder (Fig.~\ref
{fig-magnetar-2}) and compare the angular velocities at two 
points: point~$A$ where this line attaches to the pulsar and 
point~$B$ where it intersects the equator.
The angular velocity at point~$A$ is clearly equal to that of 
the neutron star: $\Omega_A = \Omega_*$. Next, what is the angular 
velocity at point~$B$? Since the field line in question extends 
beyond the light cylinder, it cannot remain purely poloidal and 
a toroidal field has to develop. Due to the assumed symmetry with 
respect to the equator, however, the toroidal field has to vanish 
at $z=0$, $B_\phi(z=0)\equiv 0$, and the plasma cannot slide 
toroidally backward along the field. Therefore, $\Omega_B$ has to 
equal the angular velocity of the plasma at this point.
Now, the toroidal field that develops in the magnetosphere off the 
equatorial plane continuously brakes the star down, so that there 
is an outward flux of angular momentum and a Poynting flux of energy 
along the field line. The rotational energy of the pulsar extracted 
by the magnetic field is partly accumulated in the magnetic form and 
partly transferred to the equatorial plasma. Thus, the material at 
point~$B$ is continuously torqued up by the magnetic field. Then, 
since the confining wall prevents the material from moving out freely 
in the radial direction, the toroidal velocity of the plasma becomes 
closer and closer to the speed of light. However, it can never exceed 
the speed of light; therefore, the angular velocity at point~$B$ is 
bounded:
$\Omega_B \simeq c/R_B = \Omega_* R_{\rm LC}/R_B$. 
Thus, we see that the field line experiences differential rotation 
at a rate $\Delta\Omega = \Omega_A-\Omega_B \geq \Omega_* (1-R_{\rm LC}/R_B)$;
For field lines with $R_B\gg R_{\rm LC}$ we get $\Delta\Omega\approx\Omega_*$. 
This differential rotation is established on the cavity's light-crossing 
time-scale, $t_0\equiv R_0/c \gg\Omega_*^{-1}$.

This differential rotation is important because it leads to 
a continuous injection of toroidal magnetic flux (of opposite 
signs) into the upper and lower hemispheres. If the cavity's 
size is fixed, then the toroidal magnetic field at any point 
grows roughly linearly with time. This is in sharp contrast 
with the unbounded pulsar case in whcih a steady state is 
established on the light travel time-scale.


Let us now analyze the field evolution on long time scales 
($t \gg t_0 \equiv R_0/c$) and at distances $R\gg R_{\rm LC}$.
Note that the toroidal magnetic field continuously increases, 
whereas the poloidal magnetic field does not. 
The poloidal electric field, $E_{\rm pol}$,  may become much larger 
than~$B_{\rm pol}$ but in any case cannot exceed the value~$B_{\rm pol} 
\Omega_* R_0 /c=B_{\rm pol} R_0/R_{\rm LC}$.
Thus, after several light-crossing times ($t\gg t_0$) the magnetosphere 
outside the light cylinder is dominated by the toroidal field, $B_\phi\gg 
E_{\rm pol}, B_{\rm pol}$. 

Next, since $B_\phi\sim t$, the relative change in~$B_\phi$ 
over~$\Delta t\sim t_0$ becomes small at late times, $t \gg t_0$. 
An approximate force-free equilibrium is then established in each 
hemisphere, described by the relativistic force-free Grad--Shafranov 
equation. In the $B_\phi$-dominated limit this equation reduces to 
$II'(\Psi)=0$, where $I\equiv R B_\phi$ is the enclosed poloidal 
current. The obvious solution of this equation is $I(\Psi)=I_0={\rm const}$,
{\it i.e.}, a singular line current $I_0(t)$ along the rotation axis. 
The toroidal magnetic field is the vacuum field produced by this
line current, $B_\phi(t,R,Z)=I_0(t)/R$; {\it i.e.}, constant on cylinders. 
The main force balance in the magnetosphere is between the toroidal 
field tension and pressure. In other words, the poloidal current 
becomes spatially separated from the toroidal magnetic field: it flows out of 
the pulsar along the axis (in both hemispheres), then as a surface current 
along the cavity walls, and finally returns to the pulsar along the 
non-force-free equatorial current sheet present due to the sharp reversal
of the toroidal field across the equator. The current density in the bulk 
of the magnetosphere is relatively small. This is similar to the electric
current structure of the magnetic bubble considered by Lyutikov \& 
Blandford\cite{lyutikov-blandford-2003}.

We shall express magnetic quantities characterizing the field in the cavity 
in terms of the total poloidal magnetic flux~$\Psi_0$ extending beyond the 
light cylinder. This flux can be crudely estimated from the pure dipole 
magnetic field, {\it i.e.},
\beq
\Psi_0 \sim \Psi_{\rm dipole}(R_{\rm LC}) = 
B_*\ {{R_*^3}\over{R_{\rm LC}}} \, .
\label{eq-Psi0}
\eeq 
Then, the characteristic poloidal magnetic field strength 
in the cavity at distances $r\sim R_0$ from the center and 
off the equatorial plane can be estimated as
\beq
B_{\rm pol} \sim B_0 \equiv {\Psi_0\over{{R_0}^2}} \sim 
B_*\, {{R_*^3}\over{R_0^2 R_{\rm LC}}} \, .
\label{eq-B0-def}
\eeq

Let us now estimate the poloidal line current~$I_0(t)$
and hence the characteristic toroidal field in the cavity. 
The poloidal current is found by following the shape of a 
field line~$\Psi$:
\beq
I(\Psi,t) = \Delta\Omega t \ \biggl[ \int\limits_\Psi 
{{dl_{\rm pol}}\over{B_{\rm pol}R^2(l_{\rm pol})}} \biggr]^{-1} \, ,
\label{eq-I-twist}
\eeq
where $l_{\rm pol}$ is the path-length along the poloidal field.
The main contribution comes from large distances, $R\sim R_0$, 
and thus, using $\Delta\Omega\simeq\Omega_*=c/R_{\rm LC}$,
we get
\beq
I_0(t) \sim \Omega_* t \ {\Psi_0\over{R_0}} \simeq
{\Psi_0\over{R_{\rm LC}}} \, {t\over t_0} \, .
\eeq
We see that for $t\gg t_0$ the poloidal current becomes much stronger 
than that in the isolated pulsar magnetosphere ($I\sim\Psi_0/R_{\rm LC}$). 
Using the estimate~(\ref{eq-Psi0}) for~$\Psi_0$, we obtain
\beq
I_0(t) \sim B_* \, {{R_*^3}\over{R_{\rm LC}^2}} \, {t\over t_0} \, .
\label{eq-I_0}
\eeq

Correspondingly, the characteristic toroidal magnetic field at distances 
of order~$R_0$ is 
\beq
B_\phi(R_0) = {I_0\over{R_0}} \simeq B_0 \Omega_* t \, ,
\label{eq-B_phi}
\eeq 
which is similar to the estimate presented by Kardashev for the toroidal 
field of a pulsar inside an expanding supernova cavity\cite{kardashev-1970}.
Thus, after many light-crossing times across the cavity, 
$B_\phi(R_0)$ becomes much larger than the toroidal field 
of an isolated pulsar at these distances 
[$B_\phi^{\rm isolated} \sim \Psi_0/(R_0 R_{\rm LC}) =
B_0 (R_0/R_{\rm LC}) = B_0 \Omega_* t_0 \ll B_0 \Omega_* t $].

\begin{figure} [h]
\centerline
{\psfig{file=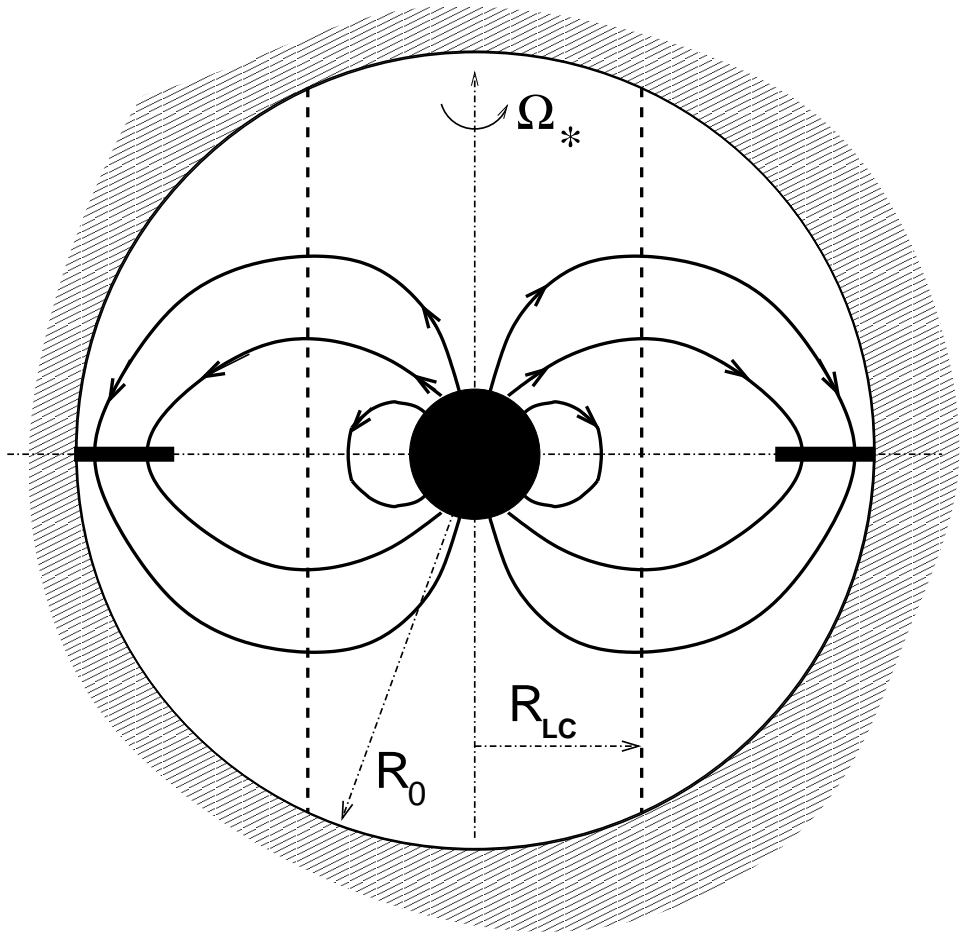,width=5in}}
\caption{At late times, the poloidal magnetic field is pressed against 
the wall by the centrifugal force of the rotating massive equatorial sheet.}
\label{fig-strip}
\end{figure}

As we noted above, the magnetosphere outside the pulsar light cylinder 
cannot be entirely force-free. Because the toroidal magnetic field 
reverses across the equator, the magnetic tension continuously accelerates 
the equatorial plasma in the toroidal direction. The tension force performs 
mechanical work on the equatorial plasma and so a part of the rotational 
energy extracted from the pulsar is deposited in the equatorial plane 
(the rest is stored magnetically in the bulk of the cavity).
Since the plasma in the equatorial plane rotates ultra-relativistically, 
the added energy leads to an increase in the relativistic ``mass'' of 
the plasma, $\Delta m\sim t^2$. Correspondingly, this relativistically
rotating massive equatorial sheet experiences an outward centrifugal 
force, $F_{\rm cent}$. 
This force cannot be balanced by the toroidal magnetic field because 
the latter is zero at the equator and so the equatorial plasma moves 
out towards the wall. It then pushes the poloidal magnetic flux out 
and concentrates it a narrow equatorial strip of ever-decreasing 
width $d(t)\ll R_0$ near the wall (see Fig.~\ref{fig-strip}).
Because of this, nearly all the poloidal flux~$\Psi_0$ that extends 
beyond the light cylinder crosses the equator at cylindrical 
radii~$R\simeq R_0$. At the same time, in the magnetosphere above and 
below the equatorial plane, the poloidal field lines that emanate from 
this strip fan out to fill the cavity volume. Thus, the characteristic 
poloidal magnetic field in the cavity is of the order~$B_0=\Psi_0/R_0^2$ 
(see eqn.~\ref{eq-B0-def}) and is much weaker (by a factor of $d/R_0$) 
than in the equatorial strip.

Let us assess the centrifugal force quantitatively.
The total torque on the massive equatorial strip is 
$\tau(t)=\int I(\Psi,t) d\Psi \simeq I_0(t)\Psi_0$
and the total work per unit time (the total Poynting
flux coming to the strip) is $P_{\rm strip}\simeq 
\tau c/R_0=B_{\phi} c\Psi_0$.
This power goes into accelerating the plasma rotation,
that is, into increasing the rotational $\gamma$-factor
and hence the relativistic mass~$m$ of the plasma in 
the strip:
\beq
m(t)c^2 = Pt \sim
\biggl({t\over{t_0}}\biggr)^2\, {{R_0}\over{R_{\rm LC}}}\,{\Psi_0^2\over{R_0}}
\sim {{R_{\rm LC}}\over{R_0}}\, B_\phi^2(t) R_0^3 \, .
\label{eq-strip-mass}
\eeq
That is, the kinetic energy in the equatorial strip is always small 
compared with the magnetic energy in the cavity, $B_\phi^2(t) R_0^3$.
The centrifugal force acting on the equatorial strip is
\beq
F_{\rm cent}(t) = {m(t)c^2\over{R_0}} \sim
B_0^2 R_0^2 \Omega_*^2 t^2\, {{R_{\rm LC}}\over{R_0}} \sim
B_\phi^2(t) R_0^2\, \biggl({{R_{\rm LC}}\over{R_0}}\biggr) \, .
\eeq
This force grows quadratically with time, just as the toroidal field 
pressure, but always remains small (by a factor of $R_{\rm LC}/R_0\ll 1$) 
compared with the total horizontal force exerted on the side wall by the 
toroidal field. However, since $F_{\rm cent}$ is concentrated in the thin 
equatorial region, it may be important in a subsequent expansion of the 
cavity (c.f. non-relativistic MHD simulations by Matt~{\it et~al.},
Ref.~\cite{matt-2006}).

A detailed analysis of the internal structure of the massive equatorial
plasma strip is an interesting problem that should be studied but it lies 
beyond the scope of this paper.


Another very important point is that the rate at which the magnetic field 
in a confined magnetosphere extracts rotational energy from the central 
rotating conductor actually grows with time. 
This is because the magnetic torque per unit area is proportional 
to the toroidal field at the conductor's surface and the latter 
grows linearly with time. Thus, as long as the cavity does not 
expand (or expands slowly) and the rotation rate of the spinning 
pulsar stays constant, the magnetic power generated by the pulsar 
inside a cavity increases linearly with time:
\beq
P(t) = I(t) \Psi_0 \Omega_*  = \Omega_*^2 t\ {\Psi_0^2\over{R_0}} \sim
P_{\rm isolated}\, {ct\over{R_0}} \, ,
\eeq
where $P_{\rm isolated}\sim B_*^2 R_*^6\Omega_*^4/c^3$ is the spin-down 
power of an isolated, unbounded pulsar. Hence, after many light-crossing 
times, the power of a pulsar-in-a-cavity greatly exceeds that of a 
classical isolated pulsar. This is resolves the apparent paradox 
raised by Lyutikov~\cite{lyutikov-2006}.

This runaway behavior can be attributed to a positive feedback 
between the energy that has been already extracted from the pulsar, 
and the strength of the agent that extracts the energy (the toroidal
magnetic field). Namely, most of the extracted energy is stored
in the toroidal magnetic field, and, since the volume is finite, 
the toroidal field strength increases with time. Since the magnetosphere 
remains in a quasi-equilibrium, the toroidal field constantly readjusts 
everywhere and the inner magnetosphere feels the presence of the outer 
confining wall. In particular, the toroidal field at the pulsar surface 
increases linearly with time, and so does the magnetic spin-down torque
on the pulsar. This picture is similar to the combustion chamber of a rocket.
In that case, the gas temperature and pressure increase as the chemical
energy of the fuel is released in the combustion process. At the same
time, the fuel burning rate grows with the ambient temperature. Therefore,
a rapid and efficient burning demands high pressure and is hence facilitated
by a strong confining chamber. Similarly, in our case of a pulsar placed
inside a cavity, the presence of strong cavity walls leads to an 
increased energy extraction rate from the pulsar.

In reality, we don't expect this power growth to last indefinitely.
It may saturate, for example, due to the development of the kink 
instability, resulting in the conversion of the toroidal flux to 
poloidal flux and to the dissipation of some of the magnetic energy 
(see Sec.~\ref{subsec-stability} for more discussion).


\subsection{Hoop-stress collimation: contrast with the isolated pulsar}
\label{subsec-collimation}

The toroidal field generated by the differential rotation exerts 
a constantly-growing pressure on the cavity walls. If we now relax 
the fixed-wall assumption, this pressure will inflate the cavity. 
Will this inflation be isotropic or, say, collimated along the axis?

Generally speaking, since the toroidal field pressure in the horizontal
direction is partly negated by the field's tension, one expects the 
resulting expansion to be mostly vertical. However, the differential 
rotation producing~$B_\phi$ is relativistic: $\Delta\Omega R_0\sim 
\Omega_* R_0\gg c$, and it is well-known that hoop-stress collimation 
is not a trivial issue in the relativistic case. Thus, it is not 
immediately obvious this mechanism can be applied to our pulsar-in-a-cavity 
scenario. 
The quintessential example of this lack of collimation for ultra-relativistic 
magnetically-dominated outflows is the isolated aligned pulsar wind inside 
the termination shock. The basic reason for this is the decollimating force 
due to the poloidal electric field, $E_{\rm pol}$. Indeed, in the case of 
an {\it unbounded} relativistic uniformly-rotating force-free magnetosphere 
in a steady state, $E_{\rm pol}$ and~$B_\phi$ are nearly equal at large 
distances from the axis\cite{gj-1969}. 
Importantly, this balance happens in an uncollimated, quasi-spherical 
poloidal magnetic field configuration; an excellent example of this is 
Michel's split-monopole solution\cite{michel-1973}. Here is a crude 
argument explaining 
this lack of hoop-stress collimation in the relativistic-rotation case.
Consider an uncollimated field configuration; the poloidal magnetic field 
is open outside the light cylinder and has a split-monopole geometry, 
{\it i.e.}, drops off as~$r^{-2}$. 
In a steady state, the poloidal electric field is $E_{\rm pol}=
B_{\rm pol}\, R/R_{\rm LC}$, where~$R$ is the cylindrical radius;
it hence drops off along radial rays as~$r^{-1}$. But $B_\phi$ 
also drops off as~$r^{-1}$. Moreover, at the light cylinder, 
$E_{\rm pol}=B_{\rm pol}\sim B_\phi$. Since outside the light 
cylinder they both decrease as~$r^{-1}$, they remain comparable 
to each other at large distances. In fact, as Goldreich \& Julian 
showed\cite{gj-1969}, $E_{\rm pol}$ and~$B_\phi$ become equal asymptotically 
as $r\rightarrow\infty$. The bottom line is that a quasi-spherical 
relativistic force-free equilibrium can be established as a balance 
between the collimating pinch force (the sum of the toroidal magnetic 
field pressure and its tension) and the opposing electric force. 
Hoop-stress collimation is suppressed as a result of this balance.

On the other hand, the case of a rotating magnetosphere enclosed 
inside a cavity is different and hoop-stress collimation can in 
fact work.  Indeed, as we showed above, at late times the toroidal
magnetic field filling the cavity becomes stronger than both~$B_{\rm
pol}$ and~$E_{\rm pol}$, in contrast to the isolated pulsar case. 
Furthermore, this toroidal field is distributed nonuniformly; namely,
$B_\phi\sim R^{-1}$. 
Correspondingly, the magnetic pressure pushing vertically against 
the top and bottom walls is much higher than that  on the side walls. 
Therefore, if we now allow the cavity to expand under this pressure, 
the expansion will be mostly vertical. The situation is then similar 
to the non-relativistic magnetic tower proposed by Lynden-Bell\cite{lb-1996}. 
We therefore envision that long-term result will be the creation of a 
pair of oppositely-directed magnetic towers\cite{um-2006a}. 
The interaction of the expanding towers with the surrounding stellar 
envelope aids in their confinement, similarly to jet collimation in 
hydrodynamical simulations of the collapsar model\cite{macfadyen-woosley-1999,
macfadyen-2001,zhang-2003}. In the scenario considered here, the towers are 
driven not by a differentially-rotating disk, but by a rapidly-rotating 
magnetar. This suggests that considering 
the pulsar magnetosphere inside a cylindrical, as opposed to spherical, 
cavity may represent yet another interesting topic for future research.

An important element in the above discussion is the fact that~$E_{\rm pol}$
is small compared with~$B_\phi$. This is because $B_\phi$ is generated as a 
result of differential rotation. This highlights the important role of {\it 
differential} rotation (as opposed to uniform relativistic rotation) in 
collimating relativistic force-free outflows.


\section{Magnetic Tower Inside a Star}
\label{sec-tower}

For simplicity, in this section we shall mostly consider the case 
when the central engine is an accretion disk around a black hole.
However, we believe that the millisecond-magnetar case is essentially
similar.



\subsection{Lynden-Bell's Original Magnetic Tower Model}
\label{subsec-lb-tower}

\begin{figure} [h]
\centerline
{\psfig{file=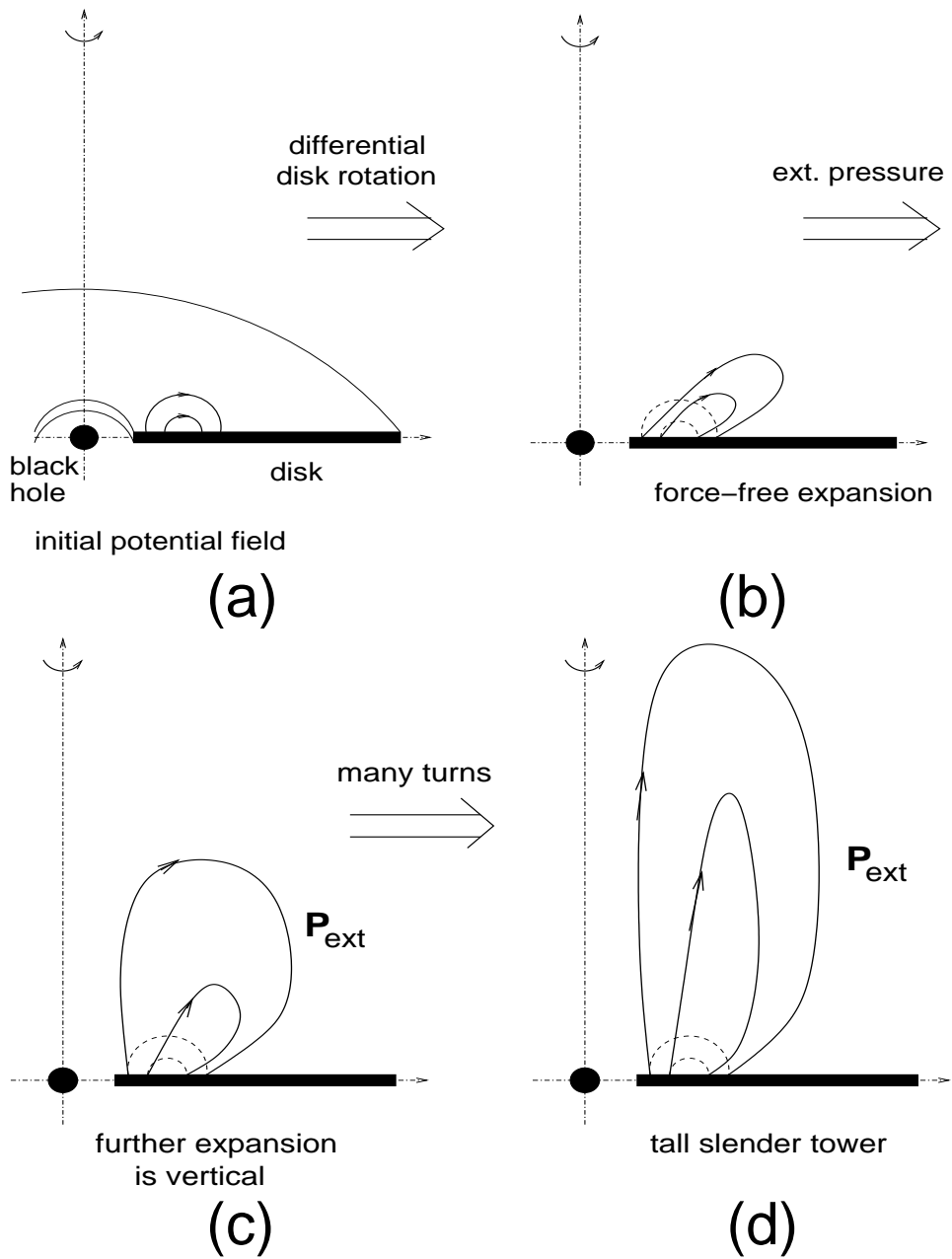,width=5in}}
\caption{Development of a magnetic tower in Lynden-Bell's (1996) model.
{\it Reproduced from Ref.~\cite{um-2006a} by permission of the AAS.}}
\label{fig-LB-tower}
\end{figure}

We suggest that the model most naturally suited to describe 
the propagation of a Poynting-flux dominated jet through a star is 
the {\it magnetic tower} model, originally introduced by Lynden-Bell 
in the AGN context\cite{lb-1996,lb-2003}.  A magnetic tower is an axisymmetric 
magnetic configuration that arises when a system of nested closed flux 
surfaces, anchored in a differentially-rotating disk, is twisted and, 
as a result, inflates, but when this inflation is controlled by a 
surrounding external pressure. The basic physical mechanism of this 
process can be described as follows (see Fig.~\ref{fig-LB-tower}).

Consider a thin conducting disk with some vertical magnetic flux frozen into
it. Let us assume that initially the magnetic field has a dipole-like topology 
(see Fig.~\ref{fig-LB-tower}a), with the
two footpoints of each field line located at different radii on the disk. 
Now let the disk rotate non-uniformly ({\it e.g.}, a Keplerian disk). 
Then, each field line~$\Psi$ is twisted at a rate $\Delta\Omega(\Psi)$ 
equal to the difference in angular velocities of its two footpoints. 
Correspondingly, toroidal magnetic flux is generated from the poloidal flux. 
The pressure of the toroidal field pushes the flux surfaces out, against the 
poloidal field tension. It is assumed that during the initial stages of this 
process the gas pressure, as well as the gravitational and inertial forces, 
are negligible in the magnetosphere, so the magnetic field is force-free. 
Then the expansion is uncollimated, typically at a $60^\circ$ angle with 
respect to the rotation axis, as shown in Figure~\ref{fig-LB-tower}b 
({\it e.g.}, Refs.~\cite{lbb-1994,ukl-2002}). However, as was shown by 
Lynden-Bell\cite{lb-1996}, if there is some, no matter how small, external
gas pressure~$P_{\rm ext}$ surrounding the expanding magnetosphere, then 
the sideways expansion ceases once $B^2/8\pi$ drops down to~$P_{\rm ext}$.
Subsequently, as again was shown by Lynden-Bell\cite{lb-1996}, the twisted 
magnetosphere continues to expand in the vertical direction
(Fig.~\ref{fig-LB-tower}c) and eventually forms a cylindrical column
that Lynden-Bell termed a {\it magnetic tower} (see Fig.~\ref{fig-LB-tower}d). 
If the external pressure outside of the tower is kept constant and uniform, 
then the top of the tower rises at a constant speed. Plasma inertia never 
plays any role; the entire evolution is a sequence of force-free magnetostatic 
equilibria with a pressure balance between the external gas outside of 
the tower and the magnetic field inside.

Note that the assumption that both ends of the field lines connect to 
the disk is not essential. A similar behavior is encountered in the case
of a rotating conducting disk magnetically connected to a central star 
({\it e.g.}, Refs.~\cite{lrbk-1995,ukl-2002}) or a rotating black hole\cite
{uzdensky-2005}. We believe that the model is also applicable to
the millisecond-magnetar central engine scenario for~GRBs.  
 
To get a physical feeling of how the magnetic tower grows, 
it is instructive to derive some simple order-of-magnitude 
estimates and scaling relationships. The main input parameters 
that set the characteristic physical scales are the total poloidal 
magnetic flux~$\Psi_0$ (per unit toroidal angle) in the tower, 
the characteristic differential rotation rate~$\Delta\Omega$, 
and the external pressure~$P_{\rm ext}$. Let us now ask how 
the main parameters of the tower, namely, its radius~$R_0$, 
the typical magnetic field~$B_0$, and the growth velocity~$V_{\rm top}$, 
scale with the three input parameters.

First, the radius of the tower, $R_0$, and the characteristic 
poloidal magnetic field, $B_{\rm pol}$ are related via 
\beq
B_{\rm pol} \sim B_0 \equiv {\Psi_0\over R_0^2} \, .
\label{eq-def-B0}
\eeq
The radius adjusts so that the magnetic pressure inside the tower 
equals~$P_{\rm ext}$. From the force-free balance inside the tower 
we expect $B_\phi\sim B_{\rm pol}\sim B_0$; hence, the total magnetic 
field strength at the outer edge of the tower is also of order~$B_0$. 
Then, from the the condition of pressure balance across the tower's 
wall we get
\beq
B_0 \sim \sqrt{8\pi P_{\rm ext}} \, , 
\label{eq-LB-side-equil}
\eeq
and so 
\beq
R_0 \sim \biggl( {\Psi_0^2\over{8\pi P_{\rm ext}}}\biggr)^{1\over4} \, .
\label{eq-LB-R0}
\eeq

To estimate the growth rate of the tower, note that the toroidal 
magnetic flux~$\chi$ is continuously generated from the poloidal 
flux~$\Psi_0$ by the differential rotation
\beq
\chi = 2\pi \Psi_0 N = \Psi_0 \Delta\Omega t   \, .
\eeq

Taking the tower to be a cylinder with radius $R_0$ and height~$Z_{\rm top}$, 
we get $B_\phi\sim \chi/R_0 Z_{\rm top}=(\Psi_0/R_0 Z_{\rm top})\,
\Delta\Omega t = B_0\, \Delta\Omega t (R_0/Z_{\rm top})$. However, 
as stated earlier, the typical toroidal field in the tower should 
be of the order of~$B_0$; therefore, the height of the tower increases 
steadily as
\beq
Z_{\rm top}(t) \sim R_0\Delta\Omega t \, .
\eeq
In other words, the tower grows at the speed of order of the typical
differential rotation velocity~$R_0\,\Delta\Omega$. If the external 
pressure does not change, the radius of the tower, determined by
Eq.~(\ref{eq-LB-R0}), stays constant during its growth; therefore, 
after many turns ($\Delta\Omega t\gg 1$), $Z_{\rm top}\gg R_0$, 
{\it i.e.}, the tower becomes slender.

Since the first analytical solution proposed by Lynden-Bell\cite{lb-1996}, 
the magnetic tower concept is becoming more and more accepted by the
astrophysical community. The formation and evolution of magnetic towers 
have been studied in numerical simulations\cite{li-2001,kato-2004a,
kato-2004b,kato-2006,nakamura-2006,nakamura-2007,ciardi-2007} and 
even in real laboratory experiments\cite{hsu-bellan-2002,lebedev-2005}.

An interesting question is the flow of energy through a magnetic tower. 
As one can easily see, Poynting flux flows up from the disk along the 
inner segment of each field line and down to the disk along the outer 
segment. Indeed, for each field line~$\Psi$, the inner, faster-rotating 
footpoint~($1$) performs work on the magnetic field, $W_1\sim I_{\rm pol}
[\Psi(1)] \Omega(1)$ (per unit time and unit poloidal flux). 
The corresponding decelerating torque per unit flux is 
$\tau_1\sim I_{\rm pol}[\Psi(1)]$. In turn, the magnetic field exerts 
an accelerating torque per unit flux $\tau_2\sim I_{\rm pol}[\Psi(2)]$ 
on the outer disk footpoint of the same field line. Correspondingly, 
it performs work at a rate $W_2\sim I_{\rm pol}[\Psi(2)]\Omega(2)$. 
Because of force-free equilibrium in the tower, $I_{\rm pol}[\Psi(1)]=
I_{\rm pol}[\Psi(2)]=I_{\rm pol}(\Psi)$, and so $\tau_1=\tau_2$, {\it i.e.}, 
all the angular momentum extracted magnetically from point~$1$ is 
transferred to point~$2$. The two energy flows, on the other hand, 
are not equal: since $\Omega(1)>\Omega(2)$, the energy extracted 
from point~$1$ along the inner segment of the field line is greater 
than the energy that flows down along the outer segment and is 
deposited in the disk at point~$2$. The difference, proportional 
to $I_{\rm pol}(\Psi)\Delta\Omega(\Psi)$, is the power driving 
the expansion of the tower. A part of it goes into filling the 
growing volume of the tower with magnetic energy, and the rest 
goes into performing work against external gas pressure and driving 
the shock through the star. 

It is interesting to note that the total vertical Poynting flux in 
the two segments only involves the differential rotation~$\Delta\Omega
=\Omega(1)-\Omega(2)$, but is independent of the absolute rotation itself.
This is because we are dealing here with a force-free equilibrium, so that 
$I_{\rm pol}$ is constant along the entire length of a field line; in 
particular, it has the same sign on the two segments of the field line,
and hence so does~$B_\phi$. The situation is drastically different
in the relativistic-rotation case where both field-line segments extend
beyond their respective light cylinders. In that case, one no longer has 
a force-free equilibrium along the entire field line; in particular, 
equilibrium breaks down at the farthermost tip of the line where the 
two segments join. As a result, the signs of~$I_{\rm pol}$ (and hence 
of~$B_\phi$) on the two segments are opposite, which corresponds to 
both segments being swept back.
Consequently, the Poynting flux is outward along both segments.
A similar situation arises in the non-force-free MHD case;
the two field-line segments are then swept back by plasma
inertia if they extend beyond the Alfv\'en point. This 
again results in an outward Poynting flux along both segments.
In both of these cases, the total vertical Poynting flux depends
on the absolute rotation rates~$\Omega(1)$ and~$\Omega(2)$ themselves, 
as opposed to just their difference.


\subsection{Magnetic Tower Driving a Shock through a Star}
\label{subsec-tower-star}

There are several reasons that make the magnetic tower 
an attractive model for the formation and propagation 
of a magnetically-dominated jet through a star within 
the collapsar model for GRBs (and core-collapse SNe). 
First, a configuration where all the field lines close back onto 
the central engine (an accretion disk or a magnetar) is natural 
for a field created by a dynamo with zero net flux. In addition, 
mixing of baryons from the stellar envelope via Kelvin-Helmholtz
and/or Rayleigh-Taylor instabilities may be inhibited by the magnetic 
field.

\begin{figure} [h]
\centerline
{\psfig{file=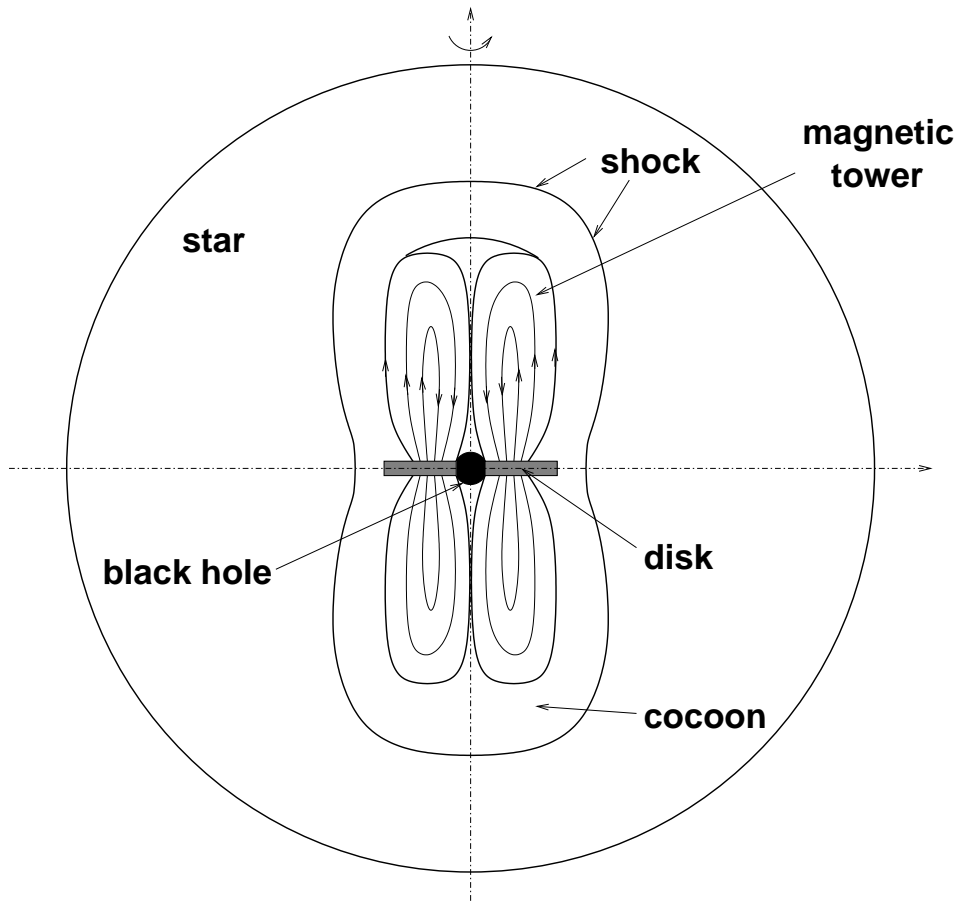,width=5in}}
\caption{Main components of a magnetic tower inside. 
The tower grows rapidly and drives a strong shock through the star. 
The shocked stellar gas behind the shock forms a hot cocoon whose 
high pressure confines the tower.
{\it Reproduced from Ref.~\cite{um-2006a} by permission of the AAS.}}
\label{fig-star}
\end{figure}

In order to apply the magnetic tower model to the collapsar scenario, we 
first want to make some modifications to Lynden-Bell's picture\cite{um-2006a}.
Specifically, we take into account the high-pressure cocoon that 
surrounds and confines the tower (see Fig.~\ref{fig-star}). 
In our model, the magnetic tower grows very rapidly and acts as 
a piston driving a shock ahead of itself. The shocked gas above 
the tower has very high pressure; it squirts sideways and forms
backflows that fill the cocoon around the tower. Therefore, the 
external pressure confining the tower is no longer an arbitrary 
parameter, as in Lynden-Bell's model, but is determined by the 
jump conditions across the shock surrounding the cocoon and across 
the contact discontinuity between the cocoon and the tower. 
The external unperturbed pressure of the star is actually irrelevant; 
it should thus be excluded from our three input parameters. Instead, 
the expansion is controlled by the ram pressure related to the gas 
{\it inertia}; therefore, we replace~$P_{\rm ext}$ by the unperturbed 
stellar density~$\rho_0$ in the list of basic dimensional parameters 
(along with $\Psi_0$ and~$\Delta\Omega$) that determine the physical 
scales in our problem. This change is an important difference between 
our model and Lynden-Bell's.

The actual situation is complicated further by the two-dimensional 
character of the problem. Since the sound speed in the cocoon is 
very high, gas pressure tends to be equalized throughout the cocoon. 
This expectation is supported by the hydrodynamic simulations of the 
collapsar model\cite{macfadyen-2001,zhang-2003}, which show a relatively 
weak (just a factor of~5 or~10) variation of the cocoon pressure along 
its length. This is very moderate compared to the corresponding variations 
of the unperturbed stellar density and pressure, which both vary by many 
orders of magnitude along the vertical extent of the cocoon. Thus, the gas 
pressure is very high everywhere in the cocoon and so the cocoon also drives 
a sideways shock into the star. The boundary between the tower and the cocoon 
is a contact discontinuity, whereas the boundary between the cocoon and the 
rest of the star is a two-dimensional strong shock of some complicated shape.


\subsection{Simple Estimates}
\label{subsec-estimates}

Let us now show how the basic parameters of the growing magnetic tower 
scale with~$\Psi_0$, $\Delta\Omega$, and~$\rho_0$. We shall ignore any 
non-uniformity of the gas pressure in the cocoon. 
Also, for simplicity we shall use one-dimensional shock jump conditions. 
Since the pressure of the unperturbed stellar gas upstream of the shock 
is neglected, the shock is strong. Assuming an adiabatic index of~5/3, 
the shock velocity with respect to the unperturbed gas is $V_s=4/3 \,
V_{\rm top}$, whereas the pressure in the post-shock region ({\it i.e.}, 
in the cocoon) is  expressed in terms of the velocity of the piston 
$V_p\equiv V_{\rm top}$ and the upstream gas density~$\rho_0$ as 
\beq
P_{\rm top} = {3\over 4}\, \rho_0 V_s^2 = {4\over 3}\,\rho_0 V_{\rm top}^2 \, .
\label{eq-shock-pressure}
\eeq

By comparing this with the pressure balance $P_{\rm top}\simeq B_0^2/8\pi$ 
at the contact discontinuity at the top of the tower, we see that the tower 
grows with a velocity of order the Alfv\'en speed corresponding to~$\rho_0$:
\beq
V_{\rm top} \sim V_A \equiv {B_0\over\sqrt{4\pi\rho_0}} =
{\Psi_0\over{R_0^2\sqrt{4\pi\rho}}}   \, .
\label{eq-V_top=V_A}
\eeq
But, as we have shown earlier, $V_{\rm top}$ should be of the order 
of~$R_0\Delta\Omega$. Thus, we obtain the scaling of~$R_0$ with~$\Psi_0$, 
$\Delta\Omega$, and~$\rho_0$:
\beq
R_0 \sim 
\biggl({\Psi_0\over{\Delta\Omega}}\biggr)^{1/3}\, (4\pi\rho_0)^{-1/6} \, .
\label{eq-R0}
\eeq
We can also relate the radius of the tower to the radius of the central 
rotating conductor. Estimating the poloidal flux as~$\Psi_0\sim B_d R_d^2$,
where $B_d$ and~$R_d$ are the typical magnetic field and the radius of the
the base of the tower ({\it e.g.}, the inner part of the accretion disk), 
we we get $B_0 \sim B_d\, (R_d/R_0)^2$ and hence
\beq
{R_0\over{R_d}} \sim \biggl({{\tilde{V}_{\rm A,d}}\over{V_d}}\biggr)^{1/3}\,.
\label{eq-R0-Rd}
\eeq
Here $V_d \equiv R_d \Delta\Omega$ is the characteristic 
differential rotation velocity of the central conductor 
and 
\beq
\tilde{V}_{\rm A,d} \equiv {B_d\over{\sqrt{4\pi\rho_0}}} 
\eeq
is a composite Alfv\'en speed involving the disk magnetic field 
and the unperturbed star's plasma density; it doesn't have a direct 
physical meaning and thus can be arbitrarily high.

Notice that, as the tower makes its way through the star, $\Psi_0$ 
and~$\Delta\Omega$ remain unchanged, whereas the third parameter, 
the unperturbed density~$\rho_0$ at the top of the tower, changes. 
It drops rather rapidly for a typical collapsar progenitor and so
the radius of the tower increases as the it grows. Thus we expect
that, in a realistic situation, the tower will not be a straight
cylinder, as we have assumed here; instead, its radius will be some
function of the vertical coordinate~$z$. 
However, as is seen from Eq.~(\ref{eq-R0}), $R_0$ scales only 
weakly with~$\rho_0$ (as $\rho_0^{-1/6}$), which somewhat justifies 
the constant-radius approximation.


\subsection{The Numbers}
\label{subsec-numbers}

Now let us make some quantitative estimates based on the above scaling 
relationships; unavoidably, these estimates will be very crude.
For definiteness, consider the accreting black hole scenario for 
the central engine.
We assume that the core of the star has collapsed into a black hole 
of fiducial mass $M=3\, M_{\odot}$ with a gravitational radius $R_g 
\equiv GM/c^2 \simeq 5\, {\rm km}$, and that some of the continuously 
infalling material has formed an accretion disk around the black hole. 
We take a fiducial disk radius of $R_d \simeq 6 R_g \simeq 3\cdot 10^6\, 
{\rm cm}$ and an initial poloidal flux (per unit toroidal angle) of~$\Psi_0= 
R_d^2 B_0 \simeq 10^{28} B_{15} R_{d,6.5}^2$ in cgs units.
This poloidal flux is being continuously twisted by the differential
rotation of the disk, with characteristic angular velocity 
$\Delta\Omega = \Omega_K(R_d) \simeq 4\cdot 10^{3} {\rm sec^{-1}} 
(M/3M_\odot)^{1/2}\, R_{d,6.5}^{-3/2}$.

Taking the fiducial stellar background density to be $\rho_0=10^6\,
{\rm g/cm^3}$, the tower outer radius can be estimated as
\beq
R_0 \sim R_d\, \biggl({{\tilde{V}_{A,d}}\over{V_d}}\biggr)^{1\over 3} =
3\, R_d\, \biggl({{B_{d,15}}
\over{R_{d,6.5}\Delta\Omega_{3.5}\sqrt{\rho_{0,6}}}}\biggr)^{1/3}\, ,
\label{eq-scaling-R_0}
\eeq
resulting in $R_0\simeq 10^7\, {\rm cm}$ for our fiducial parameter values.
Then we get the following expressions for all the other parameters:
\begin{eqnarray}
B_0 &\equiv& {\Psi_0\over{R_0^2}} = 
B_{\rm d} \biggl({R_{\rm d}\over{R_0}}\biggr)^2     \nonumber   \\
&\simeq&  0.1\, B_{\rm d}\, 
\biggl({{R_{d,6.5}\Delta\Omega_{3.5}}\over{B_{\rm d,15}}}\biggl)^{2/3}\, 
\rho_{0,6}^{1/3} \simeq 10^{14}\, {\rm G}          
\label{eq-scaling-B_0}                \, ; \\ 
V_{A,0} &\equiv& {B_0\over{\sqrt{4\pi\rho_0}}} =
3\cdot 10^{10}\, {\rm cm/sec}\ B_{d,15}^{1/3}\, R_{d,6.5}^{2/3}\,
\Delta\Omega_{3.5}^{2/3}\, \rho_{0,6}^{-1/6}    
\label{eq-scaling-V_A}            \, .
\end{eqnarray}

Notice that our crude estimate results in $V_{\rm top}\sim V_{A,0}$ 
being comparable to the speed of light~$c$. Therefore, a fully-relativistic 
treatment of the problem would be more appropriate (see Sec.~\ref{subsec-rel} 
for discussion). Such a treatment, however, lies beyond the scope of 
the current paper.

Also, we can estimate the post-shock pressure in the hot cocoon 
above the tower as
\beq
P_{\rm top} \simeq {{B_0^2}\over{8\pi}} \simeq 
4\cdot 10^{26}\, {\rm erg\ cm^{-3}}\ B_{0,14}^2 \, .
\eeq
where $B_{0,14}\equiv B_0/(10^{14}\,{\rm G})$.
At such high energy density the radiation pressure probably dominates 
over the gas pressure; we can therefore estimate the plasma temperature 
in the post-shock region as
\beq
T_{\rm top} \simeq \biggl({3P_{\rm top}\over a}\biggr)^{1/4} \simeq
2\cdot 10^{10}\, {\rm K} \simeq 2\, {\rm MeV}\, ,
\eeq
where $a\simeq 7.6\cdot 10^{-15}\, {\rm erg\, cm^{-3}\, K^{-4}}$.
On the other hand, since we are dealing with a strong hydrodynamic
shock between the cocoon and the unperturbed stellar material, the 
baryon density in the cocoon is simply $4\rho_0$, and so the 
baryon rest-mass energy density is 
$4\rho_0 c^2 \simeq 4\cdot 10^{27}{\rm erg\ cm^{-3}}\,\rho_{0,6}$,
and hence still exceeds the radiation/pair energy density by an 
order of magnitude.
The total magnetic energy contained in the tower of height~$Z_{\rm top}$
can be estimated as 
\beq
E_{\rm mag}(t) \simeq 2\pi R_0^2\, Z_{\rm top}(t)\, {{B_0^2}\over{8\pi}}
\simeq 2\cdot 10^{50}\, {\rm erg}\ R_{0,7}^2\, Z_{\rm top,9}\, B_{0,14}^2 \, ,
\label{eq-Emag}
\eeq
which is  a noticeable fraction of a typical GRB energy.


\section{Discussion}
\label{sec-discussion}


\subsection{Transition to Relativistic Expansion Regime 
and the Jet Opening Angle}
\label{subsec-rel}

As we have shown  above, the tower radius~$R_0$, and hence its growth 
velocity $V_{\rm top}\sim \Delta\Omega R_0$ scale with background density
as~$\rho_0^{-1/6}$. Therefore, as the tower expands into the outer regions 
of the star, $V_{\rm top}$ inevitably reaches the speed of light at some 
critical density~$\rho_{0,\rm rel}$.
For our fiducial values $B_d= 10^{15}\ {\rm G}$, $R_d=3\cdot10^6\ {\rm cm}$, 
and $\Delta\Omega= 3\cdot 10^3\ {\rm sec^{-1}}$, this critical density for 
the transition to the relativistic regime is $\rho_{0,\rm rel}\sim 10^6\ 
{\rm g/cm^3}$ [see eq.~(\ref{eq-scaling-V_A})]. For a typical massive 
stellar GRB progenitor this corresponds to a distance $Z_{\rm rel}$ from 
the center on the order of $10^8\ {\rm cm}$; according to Eq.~(\ref
{eq-scaling-R_0}), the corresponding radius of the tower is 
$R_{\rm rel}\sim 10^7\ {\rm cm}$. 

Our non-relativistic model becomes becomes invalid at this point and 
the subsequent expansion of the tower calls for a relativistic generalization 
of the magnetic tower model, which is beyond the scope of this paper. 
However, we probably can derive some physical insight into the relativistic
regime by looking at the results of fully-relativistic {\it hydrodynamic} 
simulations\cite{zhang-2003}. 
In those simulations the relativistic jet remained collimated as it 
propagated through the star, across several orders of magnitude in~$\rho_0$. 
This has been attributed to recollimation shocks in the cocoon and 
to relativistic beaming in the jet. But the physical processes in 
the cocoon should not change if we replace the inner relativistic 
hydrodynamic jet with a relativistic magnetic tower. Furthermore,
we expect some additional magnetic collimation due to the hoop stress.
To sum up, we expect the magnetic tower to remain collimated even after 
it transitions into the relativistic regime. In particular, we suggest 
that the final opening angle of the outflow will be about the inverse 
aspect ratio of the tower at relativistic transition:
\beq
\Delta\theta \lesssim {{R_{0,\rm rel}}\over{Z_{\rm rel}}} \simeq 0.1 \, .
\eeq
Whether this prediction is true will have to be determined by
a fully-relativistic analysis and by relativistic MHD simulations,
which we hope will be completed in the near future.


\subsection{Small-Scale Structure of the Outflow}
\label{subsec-small-scale}

\begin{figure} [h]
\centerline
{\psfig{file=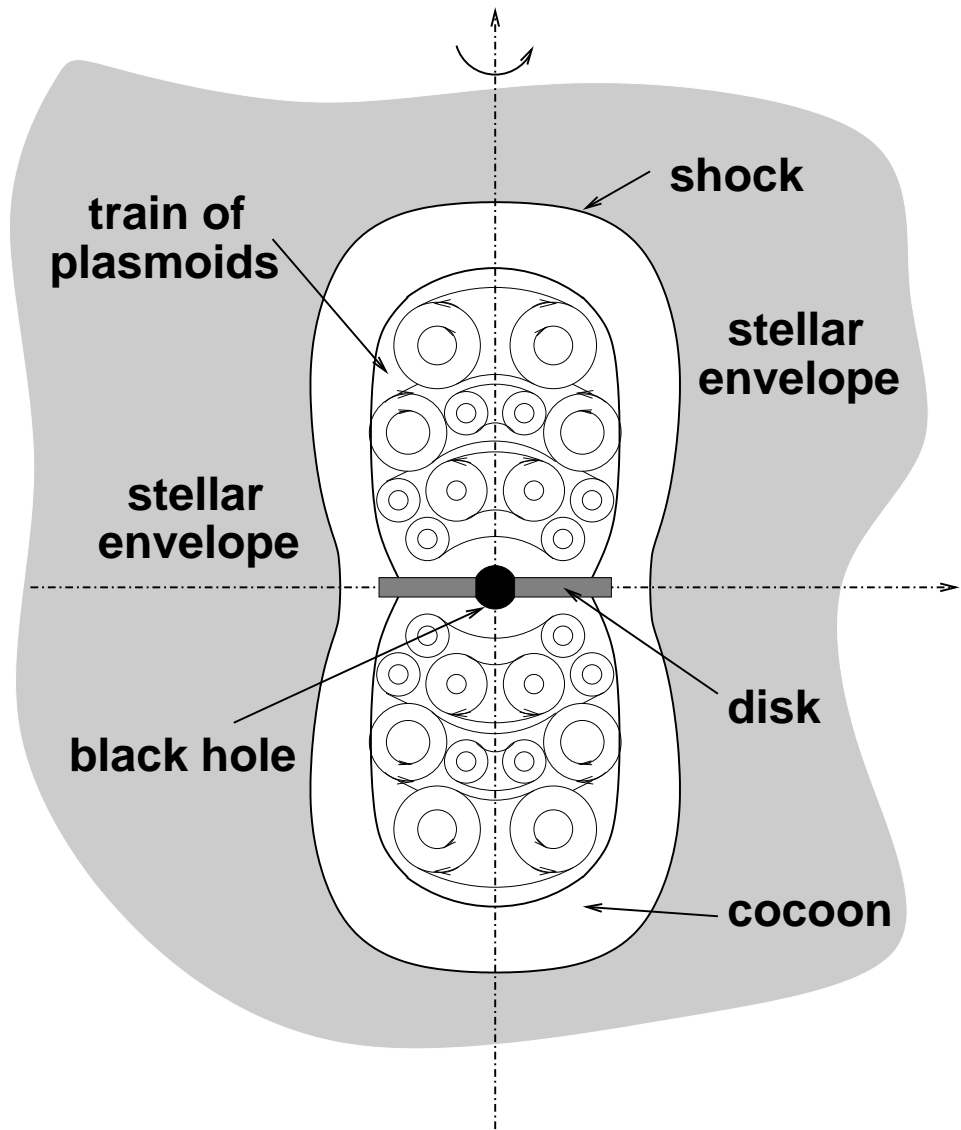,width=5in}}
\caption{Magnetic tower may have a substructure represented
by a train of many spheromak-like plasmoids. This situation may 
arise as a result of spatial and temporal intermittency at the 
base of the outflow and/or due to instabilities and reconnection
in the tower.}
\label{fig-train}
\end{figure}

The physical picture presented in this paper, with its smooth coherent
magnetic structure, is just an idealization used to get the main ideas 
across in the clearest way possible.
The actual magnetic field, especially if it is produced by a turbulent
dynamo in the PNS or in an accretion disk, will of course be different 
from such a simple axisymmetric system of nested flux surfaces. Instead, 
it may consist of an ensemble of loops of different sizes and orientations.
It may thus have a highly-intermittent substructure on smaller scales, 
both temporal and spatial. 
However, each of these smaller magnetic structures is subject to the same
physical processes as the simple large-scale configuration: twisting by
differential rotation and a subsequent inflation controlled by the external 
pressure of the cocoon and of the other loops growing at the same time. 
As a result, a more realistic picture may look like a train of spheromak-like 
plasmoids, pushing each other out along the axis (see Fig.~\ref{fig-train}). 
Hoop stress still works inside each of them, and so the overall dynamical 
effect may be qualitatively similar to that of a single tower. A similar 
picture may also develop if a large-scale twisted magnetosphere becomes 
unstable to the kink instability and undergoes flux conversion as a result, 
breaking up into smaller plasmoids (see below). In either case, the resulting 
multi-component structure of the outflow may be responsible for the observed 
intermittency in GRBs.


\subsection{Effect of MHD Instabilities on the Tower Evolution}
\label{subsec-stability}

One of greatest uncertainties in our model is the effect of 
MHD instabilities in the highly twisted magnetic structure. 
In this section we discuss two such instabilities: Rayleigh--Taylor 
and kink.

({\it i}) Rayleigh--Taylor (or its magnetic counter-part,  
Kruskal--Schwarzschild) instability may occur at the interface 
between the growing magnetic bubble and the overlying colder, 
denser stellar material. This instability is expected to affect
purely hydrodynamic fireball models as well; if anything, strong 
magnetic fields are expected to suppress it somewhat, although 
probably not completely.
The Rayleigh--Taylor instability may cause splitting of a coherent 
magnetic structure into several separate strands interlaced with 
stellar matter\cite{wheeler-2000,arons-2003}. For example, in the 
Arons model\cite{arons-2003}, the stellar envelope is ``shredded'' 
by the nonlinear Rayleigh-Taylor ``fingers''. This leads to creation 
of several evacuated channels that allow the electromagnetic energy 
produced near the central engine to escape through the star. Arons 
further argued that these channels suffer only a small amount of 
mixing with the non-relativistic stellar material due to the 
Kelvin-Helmholtz instability. In light of his work, we think that 
our magnetic cavity and/or the subsequently-formed magnetic towers 
may also suffer from fragmentation into Rayleigh-Taylor ``fingers''. 
However, as is well known, initially small-scale fingers quickly 
merge to form a few large ones in the nonlinear stage. Therefore,  
one does not expect strong mixing of the baryons from the stellar 
envelope into the magnetosphere. The exact geometry of the outflow 
may be affected somewhat and a strong time-variability may develop,
but, overall, we expect the outflow to survive. More research is needed 
in order to assess the implications of this instability for our scenario.

({\it ii}) As the magnetic configuration is twisted up, it may become 
prone to a non-axisymmetric kink-like instability. This may happen
both during the pulsar-in-a-cavity phase and during a later magnetic 
tower phase.
Whereas the stability of the pulsar-in-a-cavity has not yet been studied,
several non-relativistic 3D MHD simulations have recently addressed\cite
{kato-2004b,nakamura-2007,ciardi-2007} the kink instability of magnetic 
towers (although not in the GRB context). 
They seem to indicate that during the first few rotation periods, the tower
is stabilized by the surrounding high-pressure gas, but at later times a 
large-scale external kink does develop. As a result, the tower's general 
shape becomes helical. 
This, however, does not immediately lead to the overall disruption of 
the tower; even though the configuration is nonaxisymmetric, its main 
morphological features remain similar to those in the axisymmeitrc 
case\cite{nakamura-2007}. 
Similar conclusions have been reached by Nakamura \& Meier\cite
{nakamura-meier-2004} in their 3D-MHD study of Poynting-flux-dominated 
jets propagating through a stratified external medium. In particular, 
these authors found that a steep external pressure gradient forestalls
the instability onset. When the instability does eventually develop, 
the resulting helical structures saturate and do not develop into 
full MHD turbulence.
Important theoretical evidence supporting the idea of the external 
pressure stabilization follows from K\"onigl \& Choudhuri's analysis 
of a force-free magnetized jet confined by an external pressure\cite
{konigl-choudhuri-1985}. They showed that a non-axisymmetric helical 
equilibrium state becomes energetically favorable (conserving the total 
magnetic helicity in the jet) only when the pressure drops below a certain 
critical value. If this happens and the external kink does go unstable, 
then this non-axisymmetric equilibrium can be interpreted as the end point 
of the non-linear development of the instability.

In addition to the above non-relativistic studies, several first steps have
recently been undertaken to understand the stability of relativistic jets, 
in particular in the framework of relativistic force-free electrodynamics\cite
{gruzinov-1999,tomimatsu-2001}. However, so far as we know, there have been 
no formal stability studies of relativistic magnetic towers to date. 
Such studies, both analytical and numerical, are clearly needed. 
They would have to take into account several stabilizing effects. 
First, as Tomimatsu {\it et al.} have found in their linear stability 
analysis of a narrow rotating relativistic force-free jet, rapid 
field-line rotation inhibits the kink instability\cite{tomimatsu-2001}. 
Second, we expect that the tower expansion should quickly transition 
to the relativistic regime (see Sec.~\ref{subsec-rel}), eventually 
reaching a very large $\gamma$-factor. Once this happens, the relativistic 
time delay may effectively stabilize the tower\cite{giannios-spruit-2006}. 
This is because MHD instabilities grow on the local Alfv\'en-crossing time 
in the fluid frame and hence much slower in the laboratory frame. 
As a result, even if instabilities are excited, they do not have 
enough time to develop before the break-out of the flow from the star.

If it does develop, the kink is probably the most dangerous instability
and  may lead to a significant, although perhaps temporary, disruption. 
Such a disruption, however, is not necessarily a bad thing: the tower 
may be able to reform after being disrupted (as is seen in laboratory 
experiments\cite{lebedev-2005}) and the resulting non-steady evolution 
may provide a plausible mechanism for rapid variability seen in gamma-ray 
bursts. Also, as a result of such disruption, a significant fraction of 
the toroidal magnetic field energy may be dissipated into thermal energy\cite
{eichler-1993,begelman-1998}, which may in fact contribute to the acceleration 
of the Poynting-flux dominated outflow and to powering the prompt gamma-ray 
emission at later times\cite{drenkhahn-spruit-2002,giannios-spruit-2005,
giannios-spruit-2006,giannios-spruit-2007}. In any case, the kink, and 
especially its nonlinear outcome, is a serious issue that needs to be 
addresses in the future. Axisymmetric sausage instability also needs 
to be investigated.

An important aspect of our problem is that we are actually interested 
not so much in the instability onset or its early linear development, 
but rather in the long-term (many rotation periods) nonlinear evolution 
and its overall effect on the outflow. An important consideration that 
then needs to be taken into account is the conservation of magnetic helicity. 
Differential rotation leads to a continuous injection of helicity into the 
system (of opposite signs in the two hemispheres). In its nonlinear stage, 
the kink instability may lead to conversion of some of the toroidal magnetic 
flux to poloidal flux (in our geometry); however, it will not destroy the 
magnetic helicity accumulated in the cavity. Thus, whatever the resulting 
configuration will be, it will have to be consistent with a growing amount 
of helicity. If some of the new poloidal flux becomes detached from the base
due to reconnection (which may actually be strongly inhibited, see discussion 
in~Section~\ref{subsec-reconnection}), then the resulting configuration may
resemble a train of plasmoids (Fig.~\ref{fig-train}), similar to 
Section~\ref{subsec-small-scale}. One may in fact imagine a cyclic process 
involving the growth of the tower for several rotation periods, followed by 
flux conversion due to the kink, followed by reconnection and subsequent 
ejection of a plasmoid carrying the magnetic helicity (and some of the 
magnetic energy) injected during the given cycle.

It is also interesting to make the following observation.
An unbounded relativistic force-free outlow from a rotating conductor 
is expected to be stable. On the other hand, in the case of a confined
closed magnetosphere with field lines subject to differential rotation, 
such as our pulsar-in-a-cavity problem or a magnetic tower,  kink is 
expected to develop. At the same time, as we discussed in Sec.~\ref
{subsec-collimation}, the outflow is uncollimated in the first case
but is collimated in the second case. This suggests that there may 
be a deep connection between stability and lack of collimation of 
axisymmetric relativistic force-free flows.

Finally,  we would like to reiterate that a proper treatment of 
these problems requires a time-dependent relativistic force-free 
or full (preferably relativistic) MHD analysis and simulations 
(see~Sec.~\ref{subsec-numerical}).


\subsection{Reconnection}
\label{subsec-reconnection}

Another important issue is magnetic reconnection across the equatorial 
current sheet in the  pulsar-in-a-cavity magnetosphere or across the 
separatrix current sheet in the magnetic tower. This process may, in 
principle, lead to the break-up of a single structure into several 
smaller spheromak-like plasmoids (see Fig.~\ref{fig-train}), similar 
to the cyclic bevavior (involving reconnection) suggested for 
magnetospheres of accreting young stars\cite{goodson-1999,uzdensky-2004}. 
The expected size of the plasmoids and their production rate are 
presently not known.

We would like to remark, however, that fast a reconnection is difficult 
to achieve deep inside a collapsing star\cite{um-2006a}. 
The reason for this is that fast Petschek-like reconnection 
is now believed to be possible only in {\it collisionless} environments, 
such as the Solar corona, Earth magnetosphere, and tokamak plasmas. 
The plasma inside a collapsing star, on the other hand, is highly
collisional\cite{um-2006a}, so that classical collisional resistivity 
dominates over all other non-ideal effects in generalized Ohm's law. 
It is now believed, on the basis of numerical simulations\cite{biskamp-1986,
uzdensky-kulsrud-2000}, theoretical analysis\cite{kulsrud-2001,malyshkin-2005},
and laboratory experiments\cite{ji-1999}, that reconnection in such a 
situation proceeds in the very slow Sweet--Parker\cite{sweet-1958,parker-1957} 
regime, whereas Petschek's\cite{petschek-1964} fast mechanism fails. It is of 
course not obvious that this conclusion can be extended to the highly 
relativistic and optically-thick electron-positron plasma in the deep 
interior of a collapsing star. However, at the very least, this observation 
casts a serious doubt on the possibility of fast and efficient large-scale 
reconnection in this environment.
As the magnetic tower grows and eventually breaks out of the star, 
however, the plasma cools and the particle density in it drops rapidly. 
At some point, the plasma becomes collisionless (from the reconnection
point of view) and this opens up the possibility of reconnection and
the corresponding delayed magnetic energy release. Indeed, post-breakout 
reconnection in relativistic Poynting-flux-dominated outflows has 
been invoked as a plausible mechanism for powering GRB emission\cite
{lyutikov-blackman-2001,drenkhahn-spruit-2002,lyutikov-blandford-2003,
giannios-spruit-2005,giannios-spruit-2006,giannios-spruit-2006,lyutikov-2006}.


\subsection{Nickel Production}
\label{subsec-nickel}

A key issue for long-duration GRBs is the required production of~$^{56}$Ni.  
The supernovae observed to accompany these GRBs (SN-GRBs) belong to Type~Ibc 
\cite{soderberg-2006,kaneko-2007}. They are believed to require radioactive 
$^{56}$Ni to heat the ejecta after the initial expansion of the star. 
The brightest SN-GRBs ({\it e.g.}, SN1998bw and~SN2003dh) require up 
to several 0.1~M$_{\odot}$ of~$^{56}$Ni, as inferred from peak optical 
brightness, although on average SN-GRBs do not need more $^{56}$Ni than 
the local population of SNe\cite{soderberg-2006}. In fact, some SN-GRBs 
({\it e.g.}, GRB060505 and~GRB060614; see Refs.~\cite{fynbo-2006,
dellaValle-2006,gal-yam-2006}) may produce little or no~$^{56}$Ni
\cite{macfadyen-2003}.

In models of core-collapse SNe, $^{56}$Ni is produced in material heated to 
$T\gtrsim T_{\rm Ni} \sim 5\times 10^9$~K by the explosion shock launched in 
the core of the star.  
The amount of $^{56}$Ni produced depends on the mass inside of the expanding
shock when its temperature declines below~$T_{\rm Ni}$. This happens when its
radius has reached
\beq
R_{\rm Ni}\sim\left(\frac{3E}{4\pi a T_{\rm Ni}^4}\right)^{1/3} \sim
4 \times 10^8\, E_{51}\, {\rm cm} \, , 
\eeq
where $E=E_{51}\times 10^{51}$~erg is the explosion energy and~$a$ is 
the radiation constant. The mass inside $R_{\rm Ni}$ depends on the 
progenitor structure and on the expansion or contraction that took place
before the shock reached a given mass element. In particular, little 
$^{56}$Ni is produced by a shock, even if very powerful, if it is 
launched into a low density material, as it may occur if a weak initial 
explosion pre-inflates the stellar core so that little mass remains 
within a few $10^8$~cm when the subsequent strong shock arrives. 
Production of $\sim 0.1 M_\odot$ of $^{56}$Ni usually requires 
$\sim 10^{51}$~ergs to be deposited isotropically by a quasi-spherical 
shock within $\sim 1$~sec, so that little pre-expansion of the star occurs.
Brightest supernovae, {\it e.g.}, SN1998bw, require energies of up to
$\sim 10^{52}$~ergs to make the $\sim 0.5 M_\odot$ inferred from 
lightcurve modeling.

The requirement of fast ($\lesssim 1$~sec), isotropic deposition of energy 
for production of $^{56}$Ni is a serious challenge for models of the SN-GRB 
central engine, because the GRB engine must last for~$\gtrsim 10$~sec for 
relativistic ejecta to escape the star and because GRBs are believed to 
be collimated explosions.  
The high degree of beaming and long timescale for energy deposition makes
collapsar jets incapable of producing anywhere near the required amounts 
of~$^{56}$Ni\cite{macfadyen-woosley-1999}.  
Therefore, in the original collapsar model, with a black hole accretion disk 
as the central engine, the $^{56}$Ni is produced by a slower bi-conical disk 
wind that constitutes a distinct explosion component
\cite{macfadyen-woosley-1999,macfadyen-2003}.
In our magnetar model, on the other hand, $^{56}$Ni can be produced behind 
a roughly spherical shock driven by the initial quasi-isotropic expansion 
of the magnetosphere. 
The expansion becomes collimated and a magnetic tower forms only at 
a later stage. Thus, the collimation process involves both a quick 
isotropic expansion followed by a beamed component.  We feel that 
this modification to the magnetar scenario, {\it i.e.}, the inclusion 
of the magnetosphere interaction with the surrounding star, strengthens 
its viability as a model for the long GRB central engine.

On the other hand, one cannot exclude the possibility that, if most of 
the energy produced by the central engine escapes through the narrow jet 
channel, giving rise to a GRB, then there may not be much energy left to 
explode the rest of the star. 
If this happens, then, eventually, almost all of the stellar material 
falls into a black hole. This would include all the $^{56}$Ni that might 
have been produced, leaving no observable supernova signature 
(e.g., GRB060614).


\subsection{Pulsar Kicks}
\label{subsec-pulsar-kicks}  

Since most of the extracted rotational energy of the neutron star 
travels vertically through the two oppositely directed channels, 
a significant amount of linear momentum is also transported up and 
down and hence a reactive force is exerted on the NS from both sides
Even a small imbalance in the reactive magnetic force may impact 
a sizable overall momentum to the~NS. For example, taking the total 
initial rotational energy of the NS to be $E_{\rm rot}=5\cdot 10^{52}$~erg, 
the momentum transported out in each direction is $P=E_{\rm rot}/2c \sim 
10^{42}$~cgs (it will be even larger if the propagation speed is less 
than~$c$). Thus, just a 10\% imbalance may result in the NS terminal 
velocity of order of $v_{\rm term}\simeq 0.1 P/M_{\rm NS}\sim 300$~km/sec.


\subsection{Suggestions for Numerical Simulations}
\label{subsec-numerical}

In this section we discuss a sequence of numerical studies of the interaction 
between a magnetar's magnetosphere and its birth environment, employing a 
range of plasma descriptions. Each of them will be able to address a subset 
of key issues with increasing degree of realism. 
For example, non-relativistic axisymmetric MHD simulations can address 
the following questions:
What basic magnetic configuration results when the conditions
we describe are set up? What is the overall magnetic field structure? 
When do magnetic towers form? How does the tower shape change as it 
expands into lower-density regions?  
How strongly is the magnetic field concentrated towards the axis?  
How does the Poynting flux depend on radius and height? What is 
the effect of the neutrino- or magneto-centrifugally-driven winds? 
To what degree does the cocoon help collimate and stabilize the tower? 
How rapidly do the cocoon walls spread laterally?  Can the cocoon expansion 
result in the disruption of the star?  
In addition, some of the physical processes described in this paper may 
be relevant to other astrophysical systems, including non-relativistic 
central objects ({\it e.g.}, planetary nebulae\cite{blackman-2001,matt-2006}). 
For this reason, non-relativistic MHD simulations are of interest in 
themselves, as well as a first step toward fully-relativistic MHD. 
For example, recent non-relativistic MHD simulations indicate that 
the magnetic-tower mechanism can operate successfully in a variety 
of astrophysical environments\cite{romanova-2004,kato-2004a,kato-2004b,
nakamura-2006,matt-2006}, including collapsing massive stars\cite
{burrows-2007}.

Eventually, however, one will have to consider relativistic effects 
outside of the magnetar light cylinder or in the inner part of the black 
hole's accretion disk. This can be investigated using the relativistic 
force-free degenerate electrodynamics (FFDE) approach, valid in the case 
of a highly magnetized plasma with negligible pressure and inertia. 
Time-dependent force-free codes have recently been successfully used 
to study pulsar magnetospheres\cite{komissarov-2006,mckinney-2006, 
spitkovsky-2006}. In the pulsar-in-a-cavity context, the cavity wall 
may be represented by a rigid conducting outer boundary. While this 
case may not be directly relevant to the realistic physical environment, 
some basic aspects of a bounded rotating magnetosphere may be understood 
using this description. Furthermore, the full magnetar-in-a-star problem 
can be investigated by a hybrid simulation employing a relativistic 
force-free code inside the cavity and a relativistic hydrodynamic 
simulation outside.

Finally, relativistic-MHD simulations will be able to address questions 
fundamental to the application of magnetar-driven magnetic towers to GRBs,  
including the beaming angle and angular distribution of energy flux and 
the growth of instabilities in the relativistic outflow.
There now exist several relativistic MHD codes\cite{koide-1999, 
gammie-2003,delzanna-2003,devilliers-2003,anninos-2005,komissarov-2005,
nishikawa-2005,shibata-2006} that can be used for this problem. 
Of interest would be a set of simulations with a range of plasma~$\beta$. 
The low-$\beta$ simulations should match on to the FFDE case, at least 
qualitatively. Once these simulations are analyzed and the basic physical 
processes elucidated, $\beta$ can be gradually increased enabling an 
understanding of how plasma inertia and pressure affect the dynamics 
of the magnetosphere expansion and collimation.

The basic scenario described in this paper can initially be explored 
with axisymmetric simulations. However, in order to assess the role 
of non-axisymmetric instabilities, such as the Rayleigh--Taylor, 
Kelvin--Helmholtz, and kink instabilities (see Sec.~\ref{subsec-stability}),
three-dimensional simulations will eventually be necessary. One of the goals
of such an investigation will be to estimate and the degree of mixing of 
the envelope baryonic material into the magnetosphere. In addition, one
also would like to study the development and interaction of the MRI and 
the Parker instability. 

The operation of the Parker instability, leading to the development of 
a highly-magnetized low-density corona, may be strongly influenced by 
neutrino cooling. Thus, a realistic treatment of the neutrino heating 
and cooling processes (as it has been done, {\it e.g.}, by Burrows 
{\it et~al.}\cite{burrows-2007}) is an essential physical ingredient
of the overall problem.

In summary, numerical simulations of the full problem, including a detailed 
description of the central engine with relevant microphysical processes and 
neutrino transport, are desirable for a comprehensive understanding of the 
formation and evolution of a millisecond-magnetar-driven magnetic tower 
inside a collapsing star. We believe that such simulations will very soon
become technically feasible.


\section{Conclusions}
\label{sec-conclusions}

The core collapse of a massive rotating star may result in two distinct
outcomes, both plausible candidates for the GRB central engine.
The first one is a stellar-mass black hole with an accretion disk. 
The second is a neutron star. In this second scenario, the young 
neutron star formed as a result of core collapse has to be a millisecond 
magnetar in order to be relevant for~GRBs.

In this paper we mostly focus on the millisecond-magnetar scenario, 
although many features of our model are also applicable to the black-hole 
case, which we have considered in our previous paper (Ref.~\cite{um-2006a}). 
Of particular interest to us is the interaction between the 
rapidly-rotating magnetar's magnetosphere and the surrounding infalling 
stellar envelope. We argue that the stellar material provides a confining 
(ram) pressure that has a strong effect on both the size and the shape of 
the magnetosphere. Namely, it can channel the highly-magnetized outflow 
originating from the proto-neutron star into two collimated magnetic 
towers.

More specifically, we suggest that the stalled bounce shock --- 
a common feature in models of core-collapse supernovae --- plays 
a role of a cavity that confines the magnetosphere. 
The cavity's radius, determined by the balance between the pressure 
of the hot neutrino-heated gas and the ram pressure of the infalling 
material, stays quasi-stationary at $R_0\simeq 200$~km during the first 
few hundreds of milliseconds after the bounce.
To get a qualitative physical feeling for what happens to the magnetar 
magnetosphere during this stage, we introduce an idealized fundamental-physics 
problem that we call the {\it Pulsar-in-a-Cavity} problem\cite{um-2006b}. 
A large part of our paper (Sec.~\ref{sec-pulsar-cavity}) is devoted to 
investigating this problem.
For simplicity, we consider it under the force-free assumption.
We show that if the radius of the cavity is larger than the pulsar 
light-cylinder radius, the magnetic field inside the cavity continuously 
winds up. Then, the toroidal field strength and hence the magnetic spin-down 
luminosity of the pulsar increase, roughly linearly with time. The magnetic 
energy in the cavity grows quadratically with time. We then estimate that 
in the context of a millisecond magnetar inside a collapsing star the magnetic 
field becomes dynamically important after a few hundred turns. 
This leads to a subsequent revival of the stalled shock and may result in 
a successful magnetically-driven explosion. As long as the expansion of 
the cavity is non-relativistic, the toroidal magnetic field inside it 
remains larger than the poloidal magnetic and electric fields. As a result, 
the hoop-stress collimates the Poynting-flux-dominated outflow into two 
oppositely-directed vertical channels. 

We suggest that these magnetic outflows (either in the millisecond-magnetar 
or the accreting black hole scenario) should be described in terms of the 
magnetic tower concept, introduced by Lynden-Bell in the AGN 
context\cite{lb-1996}. 
Correspondingly, we investigate the propagation of a magnetic tower inside 
a star (see also Ref.~\cite{um-2006a}).
In our model, we modify Lynden-Bell's picture by considering that the tower 
expansion is supersonic with respect to the unperturbed stellar gas. 
We envision the growing magnetic tower acting as a piston that drives 
a strong shock through the star. The hot shocked stellar material between 
the shock and the tower forms a high-pressure cocoon that envelopes the 
tower and provides the collimating pressure. In other words, the tower 
in our model is confined not by the pressure of the background stellar 
material, but by its inertia; the strong shock and the cocoon act as 
mediators that convert the inertial support into the pressure support 
ultimately acting on the tower.
The entire configuration grows vertically with time and eventually reaches
the star's surface, thereby providing a narrow baryon-clean channel for
the Poynting-flux dominated jet, surrounded by a less-collimated hot
cocoon outflow.

Finally, we discuss the astrophysical implications of our model for 
GRBs and core-collapse supernovae, such as $^{56}$Ni production and 
pulsar kicks. In addition, we discuss the role of MHD instabilities,
most notably, the kink, in our scenario. We also assess the prospects 
for magnetic reconnection and find that it should be strongly inhibited 
in the central parts of the collapsing star owing to the high plasma 
collisionality there. Finally, we outline a set of numerical studies 
that we believe need to be done.


\acknowledgments

We would like to thank Profs.\ Jerry Ostriker and Russell Kulsrud for 
stimulating discussions and encouragement. We are also grateful to 
J.~Arons, A.~Beloborodov, E.~Blackman, P.~Goldreich, J.~Goodman, 
S.~Komissarov, A.~K\"onigl, H.~Li, M. Lyutikov, J.~McKinney, C. Thompson, 
T.~Thompson, A. Spitkovsky, H.~Spruit, J.~C. Wheeler, and E.~Zweibel 
for useful comments and suggestions. 

DAU's research has been supported by the National Science Foundation 
under Grant~PHY-0215581 (PFC: Center for Magnetic Self-Organization 
in Laboratory and Astrophysical Plasmas).
AIM acknowledges support from the Keck Fellowship at the Institute 
for Advanced Study.


\newpage




\cleardoublepage

\hoffset=2in
\voffset=1.5in

\cleardoublepage

\cleardoublepage

\cleardoublepage

\cleardoublepage

\cleardoublepage

\cleardoublepage

\cleardoublepage


\end{document}